\documentclass[a4paper, 11pt]{article}
\pdfoutput=1 

\usepackage{jheppub} 

\usepackage[T1]{fontenc} 
\usepackage{blkarray}
\usepackage{relsize}
\usepackage{booktabs}
\usepackage[bottom]{footmisc}
\usepackage{filecontents}
\usepackage[utf8]{inputenc}
\usepackage{amssymb, amsthm, tabto, epigraph, mathtools, dsfont, verbatim, slashed, mathrsfs}
\usepackage{graphicx, wrapfig}
\usepackage{enumitem}
\usepackage{relsize}
\usepackage{subcaption}
\usepackage{csquotes}
\usepackage{hyperref}
\usepackage{physics}
\usepackage{tabularx}
\usepackage{array}
\usepackage[dvipsnames]{xcolor}
\usepackage{color}
\usepackage[english]{babel}

\DeclareMathOperator{\CSgrav}{\mathrm{CS}_{\mathrm{grav}}}

\def\IZ{\mathbb{Z}}

\def\bea{\begin{eqnarray}}
\def\eea{\end{eqnarray}}
\def\beas{\begin{align*}}
\def\eeas{\end{align*}}
\def\be{\begin{equation}}
\def\ee{\end{equation}}

\usepackage{amsmath}
\usepackage{empheq}
 
\setlist{nolistsep}
\hypersetup{
    colorlinks=true,
    linkcolor=Mahogany,
    filecolor=magenta,      
    urlcolor=blue,
}

\newlength\dlf  

\allowdisplaybreaks

\setlist{nolistsep}
\hypersetup{
    colorlinks=true,
    linkcolor=Maroon,
    filecolor=Maroon,      
    urlcolor=Maroon,
    citecolor=Maroon
}

\newtheorem{theorem}{Theorem}

\newtheorem{proposition}[theorem]{Proposition}

\theoremstyle{remark}

\theoremstyle{definition}

\usepackage{amssymb}
\usepackage{tikz}
\usetikzlibrary{shapes,arrows,cd,chains,decorations.markings,decorations.pathmorphing,calc}
\tikzset{
->-/.style args={#1rotate#2}{decoration={markings, mark=at position #1 with {\arrow[scale=1.5,rotate = #2 ]{stealth}}}, postaction={decorate}}
}
\tikzset{curve/.style={settings={#1},to path={(\tikztostart)
    .. controls ($(\tikztostart)!\pv{pos}!(\tikztotarget)!\pv{height}!270:(\tikztotarget)$)
    and ($(\tikztostart)!1-\pv{pos}!(\tikztotarget)!\pv{height}!270:(\tikztotarget)$)
    .. (\tikztotarget)\tikztonodes}},
    settings/.code={\tikzset{quiver/.cd,#1}
        \def\pv##1{\pgfkeysvalueof{/tikz/quiver/##1}}},
    quiver/.cd,pos/.initial=0.35,height/.initial=0}
\tikzset{tail reversed/.code={\pgfsetarrowsstart{tikzcd to}}}
\tikzset{2tail/.code={\pgfsetarrowsstart{Implies[reversed]}}}
\tikzset{2tail reversed/.code={\pgfsetarrowsstart{Implies}}}

\newmuskip\pFqskip
\mathchardef\pFcomma=\mathcode`, 

\graphicspath{{./images/}}

\renewcommand{\bar}{\overline}

\usepackage{braket}

\usepackage[export]{adjustbox}

\title{\center{Generalized Level-Rank Duality, Holomorphic Conformal Field Theory, and Non-Invertible Anyon Condensation}}

\abstract{We study the interplay between holomorphic conformal field theory and dualities of 3D topological quantum field theories generalizing the paradigm of level-rank duality. A holomorphic conformal field theory with a Kac-Moody subalgebra implies a topological interface between Chern-Simons gauge theories. Upon condensing a suitable set of anyons, such an interface yields a duality between topological field theories.  We illustrate this idea using the $c=24$ holomorphic theories classified by Schellekens, which leads to a list of novel sporadic dualities.  Some of these dualities necessarily involve condensation of anyons with non-abelian statistics, i.e.\ gauging non-invertible one-form global symmetries. Several of the examples we discover generalize from $c=24$ to an infinite series.  This includes the fact that Spin$(n^{2})_{2}$ is dual to a twisted dihedral group gauge theory.  Finally, if $-1$ is a quadratic residue modulo $k$, we deduce the existence of a sequence of holomorphic CFTs at central charge $c=2(k-1)$ with fusion category symmetry given by $\mathrm{Spin}(k)_{2}$ or equivalently, the $\mathbb{Z}_{2}$-equivariantization of a Tambara-Yamagami fusion category.}

\author[1]{Clay C\'ordova,}
\author[1,2]{Diego Garc\'ia-Sep\'ulveda,}
\author[1]{and Jeffrey A. Harvey}

\affiliation[1]{Leinweber Institute for Theoretical Physics \& Kadanoff Center
for Theoretical Physics \& Enrico Fermi Institute, University of Chicago, Chicago, IL 60637, USA}
\affiliation[2]{Society of Fellows, Harvard University, Cambridge, MA 02138, USA}

\emailAdd{clayc@uchicago.edu}
\emailAdd{dgarciasepulveda@fas.harvard.edu}
\emailAdd{jh25@uchicago.edu}

\begin{document}

\maketitle

\section{Introduction}\label{secintro}

In this paper we explore the relationship between holomorphic 2D conformal field theory (CFT), 3D topological quantum field theory (TQFT), and anyon condensation. One primary goal is to explain how, given a 2D holomorphic CFT, one can often obtain novel dualities of 3D TQFTs related by the generalized gauging procedure of anyon condensation, i.e.\ gauging either invertible or non-invertible one-form global symmetries \cite{Gaiotto:2014kfa}. We will be particularly concerned with theories of central charge $c=24$ which we will harvest for examples, though the paradigm we explore applies to theories with general central charge.

\subsection{Holomorphic CFTs at Small Central Charge}

A rational holomorphic CFT is a vertex operator algebra (VOA).  Physically, it has purely chiral degrees of freedom.  Thus the central charges are:
\begin{equation}
    c_{L}=c~, \hspace{.2in}c_{R}=0~.
\end{equation}
The allowed values of $c$ are quantized:
\begin{equation}
    \text{bosonic:}~~~ c=8\ell ~, \hspace{.2in} \ell \in \mathbb{N}~,~~~~~  \text{spin:}~~~ c=\frac{\ell}{2} ~, \hspace{.2in} \ell \in \mathbb{N}~.
\end{equation}
Spin theories are the natural structure when the theory involves fermions.  In particular, such theories are not modular, but instead invariant under the subgroup of $SL(2,\mathbb{Z})$ that preserves a spin structure.  For recent explorations and classifications of CFTs see \cite{Creutzig:2017fuk, BoyleSmith:2023xkd, Mukhi:2022bte, Rayhaun:2023pgc, Hohn:2023auw, Moller:2024plb, Moller:2024xtt}.  Below, our primary focus is on bosonic theories that enjoy full modular invariance.

Holomorphic CFTs are often related to Kac-Moody algebras.  This is exemplified by the classification of holomorphic CFTs at small central charge:
\begin{itemize}
    \item $c=8$: There is a unique theory given by compact bosons defined on the root lattice of the Lie-Algebra $\mathrm{E}_{8}$. This is the vertex algebra $\mathrm{E}_{8,1}$ where here and in the following, the second index indicates the level of the current algebra.  The spectrum of this CFT is encoded through the torus partition function:
\begin{equation}
Z_{c=8}(\tau)=\frac{E_{4}(\tau)}{\eta(\tau)^{8}}=q^{-1/3}\left(1+248q+4124q^{2}+\cdots\right)~, \hspace{.2in}q\equiv\exp(2\pi i\tau)~. 
\end{equation}
Note in particular the leading power $q^{-c/24}$, the unique unit operator (coefficient of $q^{0}$ in parentheses), and count of the spin one conserved currents (coefficient of $q$ in parentheses).

\item $c=16$: There are two distinct theories, again both related to  Kac-Moody algebras: $\mathrm{E}_{8,1} \oplus \mathrm{E}_{8,1}$ and $\mathrm{Spin}(32)_{1}/\mathbb{Z}_{2}.$  These theories are isospectral, i.e. the set of operator dimensions and degeneracies agree and are specified by:
\begin{equation}
    Z_{c=16}(\tau)=\frac{E_{4}(\tau)^{2}}{\eta(\tau)^{16}}=q^{-2/3}\left(1+496q+69752q^{2}+\cdots\right)~. 
\end{equation}
These theories are distinguished by their operator product coefficients. 
\end{itemize}

As the central charge is increased the classification of holomorphic CFTs becomes increasingly wild.  However, there remains one case that is nearly fully understood, namely central charge $c=24,$ where the pioneering analysis was carried out by Schellekens in \cite{Schellekens:1992db}.  More specifically, \cite{Schellekens:1992db} classified the possibilities for the dimension-one part $V_1$, of a $c=24$ holomorphic VOA, and showed that there were $71$ possible choices (including the possibility that $V_{1}$ is empty). In addition, \cite{Schellekens:1992db} conjectured that each choice of $V_1$ leads to a unique VOA.  As we review below, this conjecture has essentially been proven and therefore it is again useful to organize $c=24$ holomorphic CFTs via their Kac-Moody subalgebras.\footnote{A remaining gap in the classification is to show that a holomorphic theory with $c=24$ and $V_{1}$ empty is necessarily the Monster CFT.}  For convenience, we recall this list of theories in Appendix \ref{AppSchellekens}. 

Let us now summarize the constructive progress made towards proving the conjectures of \cite{Schellekens:1992db}. The most elementary class of  $c=24$ holomorphic VOAs are lattice theories (compact bosons).  Necessarily such a lattice is positive-definite, even, and unimodular.  In rank $24$ there are $24$ such lattices classified by Niemeier: the Leech lattice, $\Lambda$, which is the unique such lattice without roots, and the $23$ Niemeier lattices $\Lambda_N$. These are uniquely specified by their roots, that is by their elements of length squared $2$. All components of the Dynkin diagrams associated with these roots have rank $24$, type A,D, or E, and have the same Coxeter number. The Niemeier lattices are constructed by adding ``glue vectors'' according to a ``glue code.'' See \cite{MR1662447} for details.

Conway, Parker, and Sloan found an elegant construction of the Niemeier lattices starting from the Leech lattice \cite{MR1662447}. They define a deep hole as a point in $\Lambda \otimes \mathbb{R}$ that has the maximal distance to the Leech lattice. Up to equivalence under the automorphism group of $\Lambda$ there are exactly $23$ deep holes and they are in one-to-one correspondence with the Niemeier lattices. Explicitly, for a deep hole $d$ the associated Niemeier lattice is the $\mathbb{Z}$-module in $\Lambda \otimes \mathbb{R}$ generated by $\{ \ell \in \Lambda| (\ell,d) \in \mathbb{Z}\}$. The construction of the Niemeier lattices from the Leech lattice leads to an analogous construction at the level of VOAs. Translation by a deep hole is an abelian automorphism of the lattice VOA $V_\Lambda$ attached to the Leech lattice of order equal to the Coxeter number of the Niemeier lattice associated to the deep hole. Gauging this automorphism, that is, constructing the orbifold theory, leads to the lattice VOA $V_N$ associated to the Niemeier lattice $N$.

The VOA $V_\Lambda$ also has automorphisms consisting of elements of the Conway group $Co_0={\rm Aut}(\Lambda)$. The center of $Co_0$ is $\mathbb{Z}_2$ and the quotient by this center is the Conway group $Co_1$, a finite, simple, sporadic group. One can construct new $c=24$ holomorphic VOAs using the orbifold construction for non-anomalous subgroups of $Co_0$. It turns out that these orbifolds exhaust the class of $c=24 $ holomorphic VOAs with $V_1 \ne \emptyset$.  See \cite{van_Ekeren_2017, Hohn:2017dsm, Hohn:2023auw, vanEkeren2021, Moller2023, hohn2022systematic, moller2024geometric} for an overview of the classification of $c=24$ VOAs in terms of what are now known as generalized deep holes of the Leech lattice.

\subsection{Holomorphic CFT, Topological Field Theory, and Interfaces}

We now elaborate on the relationship between holomorphic CFT and topological field theory that we utilize below.  

For any holomorphic CFT, the quantized value of the central charge can be interpreted as a gravitational anomaly.  By inflow, this anomaly can be understood as the level of a gravitational Chern-Simons term in 3D:
\begin{equation}\label{csgravis}
    I_{\text{inflow}}=\frac{c}{192\pi}\int \mathrm{Tr}\left(\omega \wedge d\omega+\frac{2}{3}\omega \wedge \omega \wedge \omega\right)\equiv (2c) \  \mathrm{CS}_{\text{grav}}~,
\end{equation}
where above, $\omega$ is the spin connection, and we have adopted the standard convention that a unit coefficient of gravitational anomaly corresponds to the minimal fermionic value of the chiral central charge, $c=1/2.$ The gravitational Chern-Simons term is an example of a symmetry protected topological phase (SPT) for the most fundamental symmetry of a relativistic system, Lorentz invariance.  Thus, any holomorphic CFT can be viewed as a consistent set of edge modes for this elementary SPT.

This point becomes more interesting when we examine its interplay with the Kac-Moody subalgebras of a holomorphic CFT, and recall the general discussion of \cite{Kaidi:2021gbs, Rayhaun:2023pgc}.
Suppose that the chiral algebra can be decomposed into a product of two (or more) chiral algebras of WZW form. We denote the latter factors $\mathrm{G}_{k} \times \mathrm{H}_{\tilde{k}}$, and the generalization is obvious if more than two factors are needed. The partition function of such a CFT can then be written as:
\begin{equation} \label{holomorphicpartitionfuncttion}
    Z(\tau) = \sum_{i,j} N^{ij} \chi^{\mathrm{G}_{k}}_{i}(\tau) \chi^{\mathrm{H}_{\tilde{k}}}_{j}(\tau)~,
\end{equation}
for some set of characters $\chi^{\mathrm{G}_{k}}_{i}(\tau)$ and $\chi^{\mathrm{H}_{\tilde{k}}}_{j}(\tau)$ in $\mathrm{G}_{k}$ and $\mathrm{H}_{\tilde{k}}$ respectively, and a set of nonnegative integers $N^{ij}$ specifying the pairing of characters.

\begin{figure}[t]
\centering
        \includegraphics[scale=1.8]{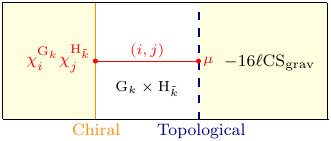} 
        \caption{The local operators of a holomorphic 2D CFT decomposed into chiral subalgebras $\mathrm{G}_{k} \times \mathrm{H}_{\tilde{k}}$ can be constructed by stretching an anyon in the $\mathrm{G}_{k} \times \mathrm{H}_{\tilde{k}}$ Chern-Simons theory from the standard chiral WZW boundary condition (orange, left) to a topological boundary condition separating $\mathrm{G}_{k} \times \mathrm{H}_{\tilde{k}}$ from an appropriate gravitational Chern-Simons term (blue, right). The index $\mu = 0,1,\ldots,N^{ij}$ specifies the pairing of the characters. The yellow background emphasizes regions with trivial anyon data.} \label{HoloCFTPartitionFunction}
\end{figure}

This result at the level of chiral algebra characters implies that the holomorphic CFT may be usefully presented from 3D as the boundary of a product Chern-Simons theory $\mathrm{G}_{k} \times \mathrm{H}_{\tilde{k}}$.  More specifically, one has Chern-Simons theory on a slab (See Figure \ref{HoloCFTPartitionFunction}).  At one end of the slab resides the non-trivial VOA realized via the standard WZW chiral boundary condition of \cite{Witten:1988hf, Elitzur:1989nr}.  Meanwhile, at the other end of the slab, the Chern-Simons theory terminates in a topological boundary, or more precisely, a topological interface to the gravitational Chern-Simons SPT \eqref{csgravis}.  Thus, the bulk boundary correspondence between VOAs and the gravitational Chern-Simons terms has been enriched, via the spin one currents of the VOA, to a thickened bulk boundary correspondence involving Chern-Simons theory.  Note also that the topological boundary restricts the endpoints of the bulk line operators to only include the subset of primaries specified in \eqref{holomorphicpartitionfuncttion}, thereby generating a suitable local operator of the holomorphic CFT from the 3D TQFT point of view.

\begin{figure}
\centering
        \includegraphics[scale=1.5]{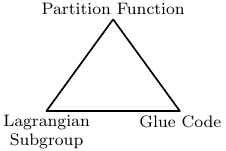} 
        \caption{Equivalence of three different concepts for Niemeier Holomorphic CFTs. The partition function can be expressed in terms of an extension of the chiral algebra, which also corresponds to gauging an abelian symmetry of the initial CFT. In 3D TQFT terms this corresponds to a Lagrangian subgroup, or gauging of an abelian one-form symmetry of the corresponding TQFT. Meanwhile, the same partition function can also be obtained from the glue code of the lattice. } \label{Triangle}
\end{figure}

One can also view the appearance of Chern-Simons theory on a slab in terms of the paradigm of symmetry TQFTs \cite{Fuchs:2002cm, Gaiotto:2014kfa, Gaiotto:2020iye, Kong_2018, Kong:2019byq, Apruzzi:2021nmk, Apruzzi:2022rei, Kaidi:2022cpf, Freed:2022qnc}. In that setup, the 2D physical theory of interest is placed at the boundary of a 3D TQFT which is the Drinfeld center of a fusion category $\mathcal{C}$.  This fusion category is a generalized (typically non-invertible) global symmetry of the boundary \cite{Chang:2018iay}. The fusion category $\mathcal{C}$ may be viewed as arising from topological lines in the bulk that are parallel to the boundary. Moreover, the precise pattern of which such lines are non-trivial at the boundary depends on the topological boundary condition terminating the slab which determines a Lagrangian algebra of the symmetry TQFT.  As constructed, any such line in $\mathcal{C}$ is not merely topological, but in fact commutes with the full Kac-Moody subalgebra of the VOA corresponding to the Chern-Simons theories. In this way presenting a holomorphic CFT via a slab of Chern-Simons theory makes manifest some of the generalized symmetries of the boundary.  See \cite{Moller:2024plb, Lin:2019hks, Fosbinder-Elkins:2024hff, Fosbinder-Elkins:2025fjz, Bae:2020pvv} for recent work on non-invertible symmetries of holomorphic CFTs.   In particular, we note that this analysis leads to an equivalence of several distinct concepts for the simple case of Niemeier holomorphic CFTs where the corresponding TQFTs and boundary fusion category symmetries are all abelian. (See Figure \ref{Triangle}.)

\subsection{Unfolding, Duality, and Anyon Condensation}

In the previous section we explained how a holomorphic CFT, together with its Kac-Moody subalgebra can be used to construct a topological boundary for the associated (product) Chern-Simons theories.  We now harness this to deduce dualities of TQFTs. 

The most important idea is simply the folding trick: a topological boundary of $\mathrm{G}_{k} \times \mathrm{H}_{\tilde{k}}$ is equivalent to a topological interface between $\mathrm{G}_{k}$ and time-reversed theory $\mathrm{H}_{-\tilde{k}}$.  Such a topological interface is, essentially, a duality of the TQFTs appearing on either side of the interface.  For instance, a line defect in $\mathrm{G}_{k}$ pierces the interface and emerges as a new line in $\mathrm{H}_{-\tilde{k}}$ thus providing the map between line defects. 

\begin{figure}[t]
\centering
        \includegraphics[scale=1.1]{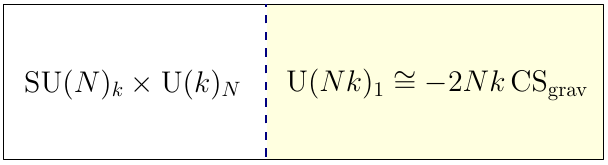} 
        \caption{A topological interface (in blue) between Chern-Simons theories $\mathrm{SU}(N)_{k} \times \mathrm{U}(k)_{N}$ and $\mathrm{U}(N k)_{1}$ follows from the existence of the conformal embedding \eqref{confembed}. The yellow background is used to emphasize that $\mathrm{U}(Nk)_{1} \cong -2 N k \CSgrav$ has trivial anyon data.} \label{BoundaryToSPT}
\end{figure}

A familiar example illustrates this construction. Consider the conformal embedding of vertex operator algebras:
\begin{equation}\label{confembed}
    \mathrm{SU}(N)_{k}\times \mathrm{U}(k)_{N}\subset \mathrm{U}( N k)_{1}~,
\end{equation}
which arises from the decomposition of $N k$, 2D complex fermions. Following the general logic of the previous section, this chiral algebra embedding yields a topological interface between the associated Chern-Simons theories \cite{Huang:2014ixa} (see also \cite{kirillov2002q,hohn2003genera}) shown in Figure \ref{BoundaryToSPT}.

\begin{figure}[t]
\centering
        \includegraphics[scale=1.1]{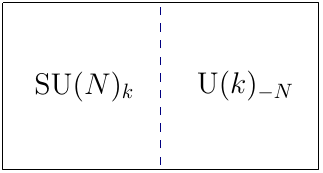} 
        \caption{A topological interface (in blue) between $\mathrm{SU}(N)_{k}$ and $\mathrm{U}(k)_{-N}$ (up to a gravitational Chern-Simons term) implies the existence of the Level-Rank duality $\mathrm{SU}(N)_{k} \cong \mathrm{U}(k)_{-N}$.} \label{LevelRankDuality}
\end{figure}

We now recall that, $\mathrm{U}(Nk)_{1}$ is in fact an invertible topological field theory.  This means that it has trivial anyon data and is equivalent to a suitable multiple of the gravitational Chern-Simons term.\footnote{We record our convention that:
\begin{equation}
    \mathrm{U}(N)_{1} \equiv -2 N \CSgrav~, \hspace{.3in} \mathrm{SO}(M)_{1} \equiv - M \CSgrav~.
\end{equation}
}
Thus, in fact the interface deduced above is a topological boundary and may be unfolded to yield an interface between $\mathrm{SU}(N)_{k}$ and $\mathrm{U}(k)_{-N}$ shown in Figure \ref{LevelRankDuality}.  This is the well-known level rank duality:
\begin{equation} \label{unitarityLRduality}
    \mathrm{SU}(N)_{k} \cong \mathrm{U}(k)_{-N}~,
\end{equation}
which holds up to a multiple of the gravitational Chern-Simons term that may be deduced from the argument above. 

Similar arguments can be used to understand the Level-Rank dualities of other gauge groups \cite{Hsin:2016blu, Aharony:2016jvv, Cordova:2017vab, Cordova:2018qvg} from the point of view of topological interfaces.

This simple example hides several features that we must confront in general below:
\begin{itemize}
    \item The conformal embedding \eqref{confembed} makes use of fermions and therefore in general only holds when the TQFTs above are viewed as spin theories. By contrast, many of the duality examples below will be valid for bosonic TQFTs.
    \item In general, a topological interface does not yield a one-to-one map of line defects across the interface.  Instead, there are a variety of possible topological junctions between the lines supported in the theories on either side of the interface.  
\end{itemize}
The second point above provides a serious challenge to the idea that topological interfaces are equivalent to dualities of TQFTs.  Fortunately a variant of this proposal is correct, but introduces a new conceptual component: \emph{anyon condensation}.

In general, anyon condensation is a gauging procedure where we modify a given TQFT $\mathcal{T}$ to a new TQFT $\mathcal{T}/\mathcal{A}$.  The quotient by $\mathcal{A}$ is formally an algebra of condensed anyons (lines), and the explicit procedure defining $\mathcal{T}/\mathcal{A}$ involves summing over suitable networks of lines in $\mathcal{A}$.  To be gaugable, the lines in $\mathcal{A}$ must, among other requirements, all be bosons and have trivial mutual statistics.  The most familiar example of anyon condensation is when $\mathcal{A}$ consists of lines with abelian fusion rules.  In that case, the lines in question define a one-form global symmetry and anyon condensation simply means gauging this one-form global symmetry.  More generally however, the lines in $\mathcal{A}$ have non-abelian statistics and non-invertible fusion rules, and the gauging operation defining $\mathcal{T}/\mathcal{A}$ is referred to as \emph{non-invertible, or non-abelian anyon condensation}.  This is the correct notion of  gauging for non-invertible one-form global symmetries in 3D TQFTs. We briefly review this procedure in Appendix \ref{GaugingReviewAppendix} and refer to \cite{Frohlich:2003hm, Bais:2008ni, Eliens:2013epa, Kong:2013aya, Burnell_2018, Cordova:2023jip, Cordova:2024goh, Kong:2024ykr, 10.21468/SciPostPhys.19.6.150, jndb-435f, Bhardwaj:2024qrf, Bhardwaj:2023bbf} for additional formalism and applications. 

In our discussion of topological interfaces and dualities of TQFTs, anyon condensation plays a central role.  As we describe more formally in Section \ref{SectionDMNO}, while a topological interface does not in general imply a duality, it does imply a duality between gauged theories.  Explicitly if $\mathcal{T}_{1}$ and $\mathcal{T}_{2}$ are TQFTs separated by a topological interface then there exist gaugable algebras $\mathcal{A}_{i}\subset \mathcal{T}_{i}$ and we have the following essential equivalence: 
\begin{equation}\label{interfaceduality}
    \mathrm{Topological \ Interface \ Between} \ \mathcal{T}_{1} \ \& \ \mathcal{T}_{2} \Longleftrightarrow \mathcal{T}_{1}/\mathcal{A}_{1} \cong \mathcal{T}_{2}/\mathcal{A}_{2}~.
\end{equation}

In the context of classical level-rank duality, such as the example presented in \eqref{unitarityLRduality}, the key statement \eqref{interfaceduality} often manifests via gauging a ``common center'' subgroup shared by both gauge groups.  This means it appears as an abelian quotient on the gauge groups. By contrast, in the most general version of level-rank duality via topological interfaces such as those discussed in \cite{Cordova:2023jip} and appearing below, non-abelian anyon condensation necessarily appears. Returning to our construction of topological interfaces via holomorphic CFTs and their Kac-Moody subalgebras, we generally obtain the duality:
\begin{equation} \label{Chiral-Antichiral-LRDuality}
    \mathrm{G}_{k}/\mathcal{A}_{\mathrm{G}} \cong \mathrm{H}_{-\tilde{k}} / \mathcal{A}_{\mathrm{H}}~,
\end{equation}
where $\mathcal{A}_{\mathrm{G}}$ and $\mathcal{A}_{\mathrm{H}}$ correspond to gauging some typically non-abelian anyons that we identify in Section \ref{SectionDMNO}.

\subsection{Example Summary}

The bulk of our analysis consists of identifying instances of the general paradigm sketched above connecting holomorphic CFTs and dualities of TQFTs.  Below we apply it often in the context $c=24$, but we also emphasize instances that extend to general central charge.

We summarize select examples here, leaving the derivations to the main text:\footnote{We use the symbol $\cong$ to mean equal up to an SPT e.g.\ a multiple of the gravitational Chern-Simons term \eqref{csgravis}. In the main text, we often explicitly specify this counterterm though we use the same notation.}

\begin{itemize}
    \item Examples that do not involve anyon condensation:
    \begin{equation}
    \mathrm{E}_{8,2} \cong \mathrm{Spin}(17)_{-1} ~, \hspace{.2in} \mathrm{E}_{7,2} \cong \mathrm{Spin}(11)_{-1} \times \mathrm{F}_{4,-1}~.
\end{equation}
\item Examples that imply time-reversal symmetries:\footnote{In Appendix \ref{SU(k)1TimeReversalAppendix} we show that $SU(k)_{1}$ is a time-reversal invariant bosonic TQFT whenever $k$ is odd and $-1$ is a quadratic residue modulo $k.$ This is the bosonic uplift of the closely related results of \cite{Delmastro:2019vnj} for spin TQFTs.}
\begin{equation}
    \mathrm{SU}(13)_{1} \cong \mathrm{SU}(13)_{-1}~, \hspace{.2in} \mathrm{SU}(7)_{1}\times \mathrm{SU}(7)_{1} \cong \mathrm{SU}(7)_{-1}\times \mathrm{SU}(7)_{-1}~,
\end{equation}
as well as:
\begin{equation}
    \frac{\mathrm{Spin(7)_{2} \times Spin(7)_{2}}}{\mathbb{Z}_{2}} \cong \frac{\mathrm{Spin(7)_{-2} \times Spin(7)_{-2}}}{\mathbb{Z}_{2}}~.
\end{equation}
These add to the existing literature of time-reversal invariant TQFTs \cite{Aharony:2016jvv, Cordova:2017kue, cheng2018microscopic, Delmastro:2019vnj, Geiko:2022qjy}.
\item Examples with abelian anyon condensation (one-form symmetry gauging):
\begin{equation}
    \frac{\mathrm{USp}(16)_{1}}{\mathbb{Z}_{2}} \cong \mathrm{F}_{4,-1} \times \mathrm{F}_{4,-1} ~, \hspace{.2in} \frac{\mathrm{E}_{6,3}}{\mathbb{Z}_{3}} \cong \mathrm{G}_{2,-1}\times \mathrm{G}_{2,-1}\times \mathrm{G}_{2,-1}~,  \hspace{.2in}\frac{\mathrm{SU}(9)_{2}}{\mathbb{Z}_{3}} \cong \mathrm{F}_{4,-2} ~,
\end{equation}
as well as the infinite sequence:
\begin{equation}
    \frac{\mathrm{SU}(2n^{2})_{1}}{\mathbb{Z}_{n}}  \cong \mathrm{E}_{7,(-1)^{n}}  \cong \mathrm{SU}(2)_{(-1)^{n+1}} ~.
\end{equation}
\item Examples involving non-abelian anyon condensation:\footnote{In the second and third examples below, we have checked consistency conditions but have not carried out a full derivation.  See section \ref{noninvertiblecondensationexamplessection}.}
\begin{equation}\label{nonabelianintro}
    \frac{\mathrm{E}_{7,3}}{\mathcal{A}} \cong \mathrm{SU}(6)_{-1}~, \hspace{.2in}\frac{\mathrm{USp}(14)_{2}}{\mathcal{A}} \cong \mathrm{SU}(4)_{-1} ~, \hspace{.2in}\frac{\mathrm{F}_{4,6}}{{\cal A}}\cong \mathrm{SU}(3)_{-2}~.
\end{equation}
as well as the following infinite pattern valid for odd $n$:
\begin{equation} \label{infinitepatternintro}
    \frac{\mathrm{Spin}(2n^{2})_{2}}{\mathcal{A}_{n}} \cong \mathrm{SU}(8)_{-1}~,
\end{equation}
where in \eqref{nonabelianintro} and $\eqref{infinitepatternintro}$ the algebras $\mathcal{A}$ and $\mathcal{A}_{n}$ are specified in section \ref{noninvertiblecondensationexamplessection}.
\item An infinite sequence of dualities:
\begin{equation}
    \mathrm{Spin}(n^{2})_{2} \cong \text{twisted Dihedral group, } \mathrm{D}_{n}, \text{ gauge theory}~,
\end{equation}
where the precise twist is specified in section \ref{sec:dihedral}.
\item Finally, given the Tambara-Yamagami fusion category\footnote{See e.g.\ \cite{Chang:2018iay, Thorngren:2021yso} for the relation of the Tambara-Yamagami fusion category to self-duality.} $TY[\mathbb{Z}_{2n+1}]$ extending the abelian group $\mathbb{Z}_{2n+1},$ we derive an equivalence between the 3D TQFT defined by its Drinfeld center $Z[TY[\mathbb{Z}_{2n+1}]]$ and a Chern-Simons theory:
\begin{equation}\label{TYintro}
    Z[TY[\mathbb{Z}_{2n+1}]]\cong \mathrm{Spin}(2n+1)_{2}\times \mathrm{SU}(2n+1)_{-1}~,
\end{equation}
reproducing observations of \cite{Ardonne_2016, Ardonne_2021, Evans:2023vns}. As we will explore in Section \ref{section5}, the relation \eqref{TYintro} explains the fact that $\mathrm{Spin}(2n+1)_{2}$ contains topological line defects that behave as self-duality defects for gauging a non-invertible global symmetry analogous to the examples of \cite{Choi:2023vgk, Diatlyk:2023fwf, Choi:2023xjw}. Using \eqref{TYintro}, we show that when $-1$ is a quadratic residue modulo $k$, $\mathrm{Spin}(k)_{2}\times \mathrm{Spin}(k)_{2}$ admits a time-reversal symmetric topological boundary. (See \cite{Benini:2018reh} for closely related discussion.)\footnote{For $k$ odd, $-1$ is a quadratic residue modulo $k$ when each prime factor is congruent to $1$ modulo $4.$} Correspondingly, this implies the existence of a sequence of holomorphic CFTs of central charge $c=2(k-1)$ with fusion category symmetry given by $\mathrm{Spin}(k)_{2}$ or equivalently, the $\mathbb{Z}_{2}$-equivariantization of $TY[\mathbb{Z}_{k}].$ 
\end{itemize}

\section{Topological Interfaces and Anyon Condensation} \label{SectionDMNO}

In this section we provide more details on the link \eqref{interfaceduality} connecting topological interfaces and dualities via anyon condensation. 

The main technical result is extracted from the literature on modular tensor categories (MTCs), which formalize the notion of a 3D TQFT.  In that context, one often makes use of the notion of Witt equivalence between MTCs (discussed below).  We have the following (Proposition 5.15) by Davydov, M\"uger, Nikshych, and Ostrik \cite{davydov2013witt}:\footnote{In fact, \cite{davydov2013witt} contains a third bullet point in Proposition 5.15, which we will, however, not utilize in this work.} 
\begin{proposition} \label{prop1}
Let $\mathcal{T}_{1}$ and $\mathcal{T}_{2}$ be MTCs. Then, the following are equivalent: \vspace{0.3cm}
\begin{itemize}
    \item $[\mathcal{T}_{1}] = [\mathcal{T}_{2}]$, i.e. $\mathcal{T}_{1}$ and $\mathcal{T}_{1}$ are Witt equivalent. \vspace{0.3cm}
    \item There exist connected \'etale algebras $\mathcal{A}_{1} \in \mathcal{T}_{1}$, $\mathcal{A}_{2} \in \mathcal{T}_{2}$ and a braided equivalence
    \begin{equation}
        \left(\mathcal{T}_{1}\right)^{0}_{\mathcal{A}_{1}} \cong \left(\mathcal{T}_{2}\right)^{0}_{\mathcal{A}_{2}}~,
    \end{equation}
\end{itemize}
\end{proposition}
\noindent where $\left(\mathcal{T}_{i}\right)^{0}_{\mathcal{A}_{i}}$ is a mathematical notation for the MTC that results after gauging the algebra 
$\mathcal{A}_{i}$ in $\mathcal{T}_{i}$. See Appendix \ref{GaugingReviewAppendix} for more precise details.
To interpret this result in physical terms, we recall (see e.g. \cite{Kaidi:2021gbs}) that MTCs $\mathcal{T}_{1}$ and $\mathcal{T}_{1}$ are Witt equivalent precisely when there exists a topological interface between the associated TQFTs. 
Meanwhile, to understand the second point above, we must recall that connected \'etale algebras describe the gauging of one-form symmetries in 3D TQFTs. See Appendix \ref{GaugingReviewAppendix} for a brief summary. In some instances, the gauging is abelian and can be interpreted in terms of modifying the global form of the gauge group (when $\mathcal{T}_{1}$ or $\mathcal{T}_{2}$ are described by some Lie group), as in the standard Level-Rank dualities \cite{Hsin:2016blu, Aharony:2016jvv, Cordova:2017vab, Cordova:2018qvg}. Importantly, however, this is not the most general scenario, and the gauging above may involve non-abelian anyons. When this is the case, we do not have an interpretation as simple as modifying the global form of the gauge group, but as we will see below in concrete examples (see also \cite{Cordova:2023jip}), this non-abelian condensation can be made fully explicit. This physical interpretation of Proposition \ref{prop1} justifies \eqref{interfaceduality}.

\subsection{Identification of the Lagrangian Algebra and Condensing Anyons}

\begin{figure}[t]
\centering
        \includegraphics[scale=1.5]{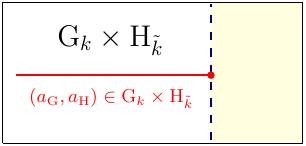} 
        \caption{The anyons in the $\mathrm{G}_{k} \times \mathrm{H}_{\tilde{k}}$ decomposition of the holomorphic chiral algebra define a Lagrangian algebra in the $\mathrm{G}_{k} \times \mathrm{H}_{\tilde{k}}$ Chern-Simons theory.  This specifies the lines that can terminate on the topological boundary. The yellow background emphasizes the region with trivial anyon data.} \label{LagrangiantoFrobenius}
\end{figure}

Our discussion above has rigorously justified the idea that holomorphic CFTs with Kac-Moody subalgebras leads to topological interfaces between Chern-Simons theories and corresponding dualities upon anyon condensation:
\begin{equation}\label{dualitymain}
    \mathrm{G}_{k}/\mathcal{A}_{\mathrm{G}} \cong \mathrm{H}_{-\tilde{k}} / \mathcal{A}_{\mathrm{H}}~,
\end{equation}
for some condensable algebras $\mathcal{A}_{\mathrm{G}}$ and $\mathcal{A}_{\mathrm{H}}$ (and up to a suitable multiple of the gravitational Chern-Simons counterterm).  To complete our discussion, it remains to identify the algebras $\mathcal{A}_{\mathrm{G}}$ and $\mathcal{A}_{\mathrm{H}}$.

To carry this out we proceed as follows.  First, note that before unfolding, the topological boundary of $\mathrm{G}_{k}\times \mathrm{H}_{\tilde{k}}$ may itself be characterized in terms of a set of gauged anyons known as a Lagrangian algebra \cite{davydov2013witt, Kong:2013aya, Cong:2017ffh, Kaidi:2021gbs}.  This is a maximal set of gaugable anyons such that after condensation the resulting theory becomes trivial (i.e.\ invertible).  In particular, the lines  $(a_{\mathrm{G}},a_{\mathrm{H}}) \in \mathrm{G}_{k} \times \mathrm{H}_{\tilde{k}}$ that are in the Lagrangian algebra are precisely those that can end at the topological boundary, as depicted in Figure \ref{LagrangiantoFrobenius}. In terms of the character decomposition \eqref{holomorphicpartitionfuncttion}, these pairs are precisely those that appear in the sum defining the partition function.  

\begin{figure}
\centering
        \includegraphics[scale=1.5]{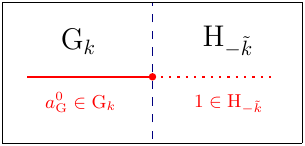} 
        \caption{Unfolding, the Lagrangian algebra gives rise to a certain set of anyons $a_{\mathrm{G}}^{0} \in \mathrm{G}_{k}$ that, by definition, couple with the identity of $\mathrm{H}_{-\tilde{k}}$ on the right-hand side. We propose this set of anyons generate the Frobenius algebra $\mathcal{A}_{\mathrm{G}}$ in $\mathrm{G}_{k}$ appearing in the resulting Level-Rank duality. An analogous observation allows us to construct the Frobenius algebra $\mathcal{A}_{\mathrm{H}}$ in $\mathrm{H}_{-\tilde{k}}$.} \label{ReadingFrobeniusAlgebrafromLagrangian}
\end{figure}

Upon unfolding, the pair $(a_{\mathrm{G}},a_{\mathrm{H}})$ implies the existence of a topological junction between the lines $a_{\mathrm{G}}$ in $\mathrm{G}_{k}$ on one side of the interface and $a_{\mathrm{H}}$ in $\mathrm{H}_{-\tilde{k}}$ on the other side. In particular to extract the algebras appearing in the duality statement \eqref{dualitymain} we set $a_{\mathrm{H}}=1$ and determine all the lines $a^{0}_{\mathrm{G}}$ that have a topological junction with the identity line in $\mathrm{H}_{-\tilde{k}}$. Then, the anyons that generate the condensable algebra in $\mathrm{G}_{k}$ appearing in the duality are precisely this set of anyons with a junction to the identity line (see Figure \ref{ReadingFrobeniusAlgebrafromLagrangian}): 
\begin{equation}\label{Frobenius}
    \mathcal{A}_{\mathrm{G}} = \sum a^{0}_{\mathrm{G}}.
\end{equation}
Similar remarks hold for $\mathrm{H}_{-\tilde{k}}$. We will see a variety of examples of this identification below. Note that in the most general situation both $\mathcal{A}_{\mathrm{G}}$ and $\mathcal{A}_{\mathrm{H}}$ can be non-trivial simultaneously.\footnote{For example, the duality $\mathrm{PSU(3)_{6} \cong SO(3)_{-8}}$ arising from the conformal embedding  $\mathrm{SU(3)_{6} \times SU(2)_{16} \subset \mathrm{E}_{8,1}}$ requires condensation in both theories \cite{Cordova:2018qvg}.}  However, in the specific instances we study in this paper, at most one of the algebras will be non-trivial.

\subsection{Comments on Time-Reversal}

One class of examples of particular interest are dualities between a theory $\mathcal{T}$ and its time-reversal $\overline{\mathcal{T}}.$  Any such duality leads to a time-reversal symmetry of the theory $\mathcal{T}$.

To connect to the previous discussion, if $\mathcal{T}$ enjoys time-reversal invariance, it necessarily entails the existence of a topological interface separating $\mathcal{T}$ and its time-reversal $\overline{\mathcal{T}}$. However, from the discussion below Proposition \ref{prop1} we know that in general such an interface implies that time-reversal invariance holds only up to a condensation in $\mathcal{T}$:
\begin{equation}
    \mathcal{T}/\mathcal{A} \cong \overline{\mathcal{T}/\mathcal{A}}~.
\end{equation}
This result can be improved if one uses our observations about the Lagrangian algebra around \eqref{Frobenius}.   Namely, if after unfolding the only line with a junction to the identity line in $\overline{\mathcal{T}}$ is the identity line in $\mathcal{T}$ (and vice versa), then the algebras trivialize on both sides of the relation, and one finds: 
\begin{equation}\label{Tdual}
    \mathcal{T} \cong \overline{\mathcal{T}}.
\end{equation}

We can also derive \eqref{Tdual} rigorously from the coset inversion theorem of \cite{Frohlich:2003hm} (see also the discussion in \cite{Cordova:2023jip, Cordova:2025eim}). Specifically, if one has MTCs $\mathcal{T}_{1}$, $\mathcal{T}_{2}$ and $\mathcal{T}_{3}$ such that
\begin{equation}\label{cinv}
    \mathcal{T}_{1} = \frac{\mathcal{T}_{2} \times \mathcal{T}_{3}}{\mathcal{A}_{1}}
\end{equation}
for some algebra $\mathcal{A}_{1}$ in the MTC $\mathcal{T}_{2} \times \mathcal{T}_{3}$, then as long as the only element of the form $(\alpha_{2}, 1)$ or $(1, \beta_{3})$ in $\mathcal{T}_{2} \times \mathcal{T}_{3}$ is the identity line $(1,1)$, one can ``solve'' \eqref{cinv} for either $\mathcal{T}_{2}$ or $\mathcal{T}_{3}$. Explicitly:
\begin{equation}
    \mathcal{T}_{2} = \frac{\mathcal{T}_{1} \times \overline{\mathcal{T}_{3}}}{\mathcal{A}_{2}}~, \quad  \quad \mathcal{T}_{3} = \frac{\mathcal{T}_{1} \times \overline{\mathcal{T}_{2}}}{\mathcal{A}_{3}}~,
\end{equation}
for some algebras $\mathcal{A}_{2}, \mathcal{A}_{3}$ in $\mathcal{T}_{1} \times \overline{\mathcal{T}_{3}}$ and $\mathcal{T}_{1} \times \overline{\mathcal{T}_{2}}$ respectively.

Consider now applying the coset inversion formula with $\mathcal{T}_{1}$ the trivial MTC, and $\mathcal{T}_{2} = \mathcal{T}_{3} = \mathcal{T}$. This matches our setup where a product of two theories has a boundary, and we interpret $\mathcal{A}_{1}$ as the Lagrangian defining the boundary condition.  Using the previously mentioned assumption on the algebra $\mathcal{A}_{1}$, we obtain:
\begin{equation}
     \mathcal{T} = \frac{\overline{\mathcal{T}}}{\mathcal{A}_{2}}=\frac{\overline{\mathcal{T}}}{\mathcal{A}_{3}}~.
\end{equation}
Upon taking quantum dimensions of both sides (see \eqref{quantumdimensionconstraint}) we find:
\begin{equation}
    \mathrm{dim}\left(\mathcal{T}\right)=\frac{\mathrm{dim}\left(\overline{\mathcal{T}}\right)}{\mathrm{dim}\left(\mathcal{A}_{2}\right)^{2}}=\frac{\mathrm{dim}\left(\overline{\mathcal{T}}\right)}{\mathrm{dim}\left(\mathcal{A}_{3}\right)^{2}} ~.
\end{equation}
Since the quantum dimensions of $\mathcal{T}$ and $\overline{\mathcal{T}}$ are equal, this impies that the algebras $\mathcal{A}_{2}$ and $\mathcal{A}_{3}$ are trivial.  Thus:
\begin{equation}
    \mathcal{T} \cong \overline{\mathcal{T}}~,
\end{equation}
which is indeed the result claimed above.

\section{Examples}

In this section we consider explicit examples of duality obtained by constructing topological interfaces from holomorphic CFTs. We begin with examples that require no gauging by either abelian or non-abelian anyons. A special kind of example of this type corresponds to time-reversal symmetric TQFTs, which we consider separately. Then, we consider examples involving abelian gauging which have the interpretation of modifying the global form of the gauge group into an appropriate non-simply-connected version. Finally, we consider examples that require gauging of non-invertible one-form symmetries (non-abelian anyon condensation). The spectrum and fusion rules of the MTCs used in this section are obtained from the KAC software program \cite{KAC}.

\subsection{Examples without Condensation}

\begin{table}[t]
\centering
\begin{tabular}[h]{|p{2cm}|p{3cm}|p{2.5cm}|p{2.5cm}| }
\hline 
\multicolumn{4}{|c|}{$\mathrm{E}_{8,2}$ } \\
\hline
Line label & Highest weight & Quantum Dim. & Conf. Weight \\
\hline
0 & 0  & $d_{0} = 1$ & $h_{0} = 0$ \\
1 & $\mathbf{w}_{7}$ & $d_{1} = 1$ & $h_{1} = 3/2$ \\
2 & $\mathbf{w}_{1}$  & $d_{2} = \sqrt{2}$ & $h_{2} = 15/16$ \\
\hline
\end{tabular} 
\caption{Data of $\mathrm{E}_{8,2}$.}  \label{E8lv2table}
\end{table}

We begin with some examples that do not require gauging of any one-form symmetry. A case, readily apparent upon direct examination, is taken from \cite{Schellekens:1992db} number 62 (see Appendix \ref{AppSchellekens}). Following our previous discussion, and using the spectra of line operators shown in Tables \ref{E8lv2table} and Appendix \ref{SummaryAnyonDataAppendix}, it is straightforward to verify that:
\begin{equation}
    \mathrm{E}_{8,2} \cong \mathrm{Spin}(17)_{-1} - 48 \CSgrav.
\end{equation}
Meanwhile, if we fermionize (extending by the current of spin $3/2$ on the left-hand side, while extending respect to the $1/2$ vectorial line on the right-hand side) we obtain that, as a spin theory, $\mathrm{E}_{8,2}$ is a purely gravitational Chern-Simons term:
\begin{equation}
    \mathrm{E}_{8,2} \cong - 31 \CSgrav.
\end{equation}

A second, less trivial example that does not involve condensation can be obtained from \cite{Schellekens:1992db} element number 53 (see Appendix \ref{AppSchellekens}). Specifically:
\begin{equation}
    \mathrm{E}_{7,2} \cong \mathrm{Spin}(11)_{-1} \times \mathrm{F}_{4,-1} - 48 \CSgrav,
\end{equation}
which again is easy to verify using the spectra of line operators shown in Table \ref{E7lv2table} and Appendix \ref{SummaryAnyonDataAppendix}. Similarly as in the previous example, we can fermionize both sides, extending by the $3/2$ spin primary on the left-hand side and extending by the vectorial, spin $1/2$ line on the right-hand side. We obtain the spin Level-Rank duality:
\begin{equation}
    \mathrm{E}_{7,2} \cong \mathrm{F}_{4,-1} - 37 \CSgrav.
\end{equation}

\begin{table}[t]
\centering
\begin{tabular}[h]{|p{2cm}|p{2.5cm}|p{3cm}|p{2.5cm}| }
\hline 
\multicolumn{4}{|c|}{$\mathrm{E}_{7,2}$ } \\
\hline
Line label & Highest weight & Quantum Dim. & Conf. Weight \\
\hline
0 & 0  & $d_{0} = 1$ & $h_{0} = 0$ \\
1 & 2$\mathbf{w}_{6}$ & $d_{1} = 1$ & $h_{1} = 3/2$ \\
2 & $\mathbf{w}_{7}$  & $d_{2} = \sqrt{2}$ & $h_{2} = 21/16$ \\
3 & $\mathbf{w}_{6}$ & $d_{3} = (1+\sqrt{5})/\sqrt{2}$ & $h_{3} = 57/80$ \\
4 & $\mathbf{w}_{5}$ & $d_{4} = (1 + \sqrt{5})/2$ & $h_{4} = 7/5$ \\
5 & $\mathbf{w}_{1}$ & $d_{5} = (1 + \sqrt{5})/2$ & $h_{5} = 9/10$ \\
\hline
\end{tabular} 
\caption{Data of $\mathrm{E}_{7,2}$.}  \label{E7lv2table}
\end{table}

\subsection{Time-Reversal Invariant Examples}

Several examples of bosonic time-reversal invariance follow from \cite{Schellekens:1992db}:
\begin{equation}
    \mathrm{SU(13)_{1} \cong SU(13)_{-1}},
\end{equation}
Tensoring by a transparent fermion (i.e $\mathrm{SO}(0)_{1}$), by the level-rank duality $\mathrm{SU}(N)_{k} \cong \mathrm{U}(k)_{-N}$ we reproduce the time-reversal of $U(1)_{13}$ classified in \cite{Delmastro:2019vnj}. Several other examples follow from \cite{Schellekens:1992db}, generalizing the argument in \cite{Aharony:2016jvv} to find time-reversal invariant TQFTs from conformal embeddings. We mention a few interesting examples:
\begin{equation}
    \mathrm{SU(7)^{2}_{1} \cong SU(7)^{2}_{-1}},
\end{equation}

\begin{equation}
    \frac{\mathrm{Spin(7)_{2} \times Spin(7)_{2}}}{\mathbb{Z}_{2}} \cong \frac{\mathrm{Spin(7)_{-2} \times Spin(7)_{-2}}}{\mathbb{Z}_{2}}.
\end{equation}

Of course, any of these examples can be lifted to a duality of spin theories by tensoring with the transparent fermion. Further examples for spin theories that cannot be obtained this way can be found from the classification of fermionic chiral CFTs, see e.g. \cite{BoyleSmith:2023xkd, Rayhaun:2023pgc}. From this we find, for example
\begin{equation}
    \frac{\mathrm{SU(3)_{1} \times SU(6)_{2}}}{\mathbb{Z}_{2}} \cong \frac{\mathrm{SU(3)_{-1} \times SU(6)_{-2}}}{\mathbb{Z}_{2}},
\end{equation}
where the gauging is by a fermionic current. Similarly:
\begin{equation}
    \frac{\mathrm{USp}(12)_{1}}{\mathbb{Z}_{2}} \cong \frac{\mathrm{USp}(12)_{-1}}{\mathbb{Z}_{2}}.
\end{equation}
Additional examples may be constructed in a similar manner.

\subsection{Invertible Condensation Examples}

In this subsection we showcase examples where it is necessary to gauge some abelian one-form symmetry. The dualities are straightforwardly verified by making of the three-step gauging rule \cite{Moore:1989yh,Hsin:2018vcg}, which we recall in Appendix \ref{GaugingReviewAppendix}.

\subsubsection*{$\mathrm{A}_{17,1} \times \mathrm{E}_{7,1}$}

\begin{table}[t]
\centering
\begin{tabular}[h]{|p{2cm}|p{2.5cm}|p{3cm}|p{2.5cm}| }
\hline 
\multicolumn{4}{|c|}{$\mathrm{E}_{7,1}$ } \\
\hline
Line label & Highest weight & Quantum Dim. & Conf. Weight \\
\hline
0 & 0  & $d_{0} = 1$ & $h_{0} = 0$ \\
1 & $\mathbf{w}_{6}$ & $d_{1} = 1$ & $h_{1} = 3/4$ \\
\hline
\end{tabular} 
\caption{Data of $\mathrm{E}_{7,1}$.}  \label{E7lv1table}
\end{table}

We begin with an easy-to-verify example involving purely abelian MTCs. This is:
\begin{equation} \label{SomeLevelRankDuality}
    \frac{\mathrm{SU}(18)_{1}}{\mathbb{Z}_{3}} \cong \mathrm{E}_{7,-1} - 48 \CSgrav ,
\end{equation}
corresponding to \cite{Schellekens:1992db} entry number 65 (see Appendix \ref{AppSchellekens}). To verify this, we can read the spectrum of $\mathrm{SU}(18)_{1}$ from Appendix \ref{SummaryAnyonDataAppendix}, apply the three-step procedure recalled in \ref{GaugingReviewAppendix}, and compare with the spectrum of $\mathrm{E}_{7,1}$ presented in Table \ref{E7lv1table}. The rank-six and rank-twelve antisymmetric representations have integer topological spin, and as such we can gauge the $\mathbb{Z}_{3}$ symmetry furnished by these lines. Then, the only gauge-invariant lines are the ones in the rank-$3n$ antisymmetric representations with $n = 0,1,2,3,4,5$. Because of the gauge symmetry, the lines $n=0,2,4$ are identified with each other, and similarly for $n=1,3,5$. We find, as expected, that there are only two remaining lines that precisely match the spectrum of $\mathrm{E}_{7,-1}$.

Moreover, from the study of conformal embeddings, one is able to derive the Level-Rank dualities $\mathrm{E}_{7,1} \cong \mathrm{SU}(2)_{-1} - 16 \CSgrav \cong \mathrm{SU}(8)_{1}/\mathbb{Z}_{2}$ (See \cite{Cordova:2018qvg}). Thus, we can add the duality \eqref{SomeLevelRankDuality} to this sequence of Level-Rank dualities, and write:
\begin{equation} \label{ExceptionalLevelRankDuality}
    \frac{\mathrm{SU}(18)_{1}}{\mathbb{Z}_{3}} \cong \mathrm{E}_{7,-1} - 48 \CSgrav \cong \mathrm{SU}(2)_{1} - 32 \CSgrav \cong \frac{\mathrm{SU}(8)_{-1}}{\mathbb{Z}_{2}} - 48 \CSgrav
\end{equation}
In fact, performing the gauging explicitly, we may derive
\begin{equation}
    \frac{\mathrm{SU}(2n^{2})_{1}}{\mathbb{Z}_{n}} + 2(2n^{2} - 1) \CSgrav \cong \mathrm{E}_{7,(-1)^{n}} + (-1)^{n} 14 \CSgrav \cong \mathrm{SU}(2)_{(-1)^{n+1}} + (-1)^{n+1} 2 \CSgrav,
\end{equation}
which generalizes the pattern \eqref{ExceptionalLevelRankDuality} to arbitrary nonnegative integer $n$.

\subsubsection*{$\mathrm{C}_{8,1} \times (\mathrm{F}_{4,1})^{2}$}

Our next example corresponds to a Level-Rank duality involving abelian gauging of non-abelian MTCs. More precisely, we take element number 52 \cite{Schellekens:1992db} (see Appendix \ref{AppSchellekens}). The spectrum of $\mathrm{USp}(16)_{1}$ is shown in Table \ref{USp16lv1table}, and that of $\mathrm{F}_{4.1}$ is summarized in Appendix \ref{SummaryAnyonDataAppendix}. We obtain:
\begin{equation}
    \frac{\mathrm{USp}(16)_{1}}{\mathbb{Z}_{2}} \cong \mathrm{F}_{4,-1} \times \mathrm{F}_{4,-1} - 48 \CSgrav
\end{equation}
Following the three-step procedure (see Appendix \ref{GaugingReviewAppendix}), we see that the lines labeled $\mathbf{w}_{7}$, $\mathbf{w}_{1}$, $\mathbf{w}_{5}$, $\mathbf{w}_{3}$ are not gauge invariant under the $\mathbb{Z}_{2}$ center symmetry (corresponding to the line of highest-weight $\mathbf{w}_{8}$), and thus are removed from the spectrum. Meanwhile, $\mathbf{w}_{6}, \mathbf{w}_{2}$ identify with each other, and the line $\mathbf{w}_{4}$ splits into two lines, indeed recovering properly the spectrum of $\mathrm{F}_{4,-1} \times \mathrm{F}_{4,-1}$ with correct quantum dimensions and topological spins.

\begin{table}[t]
\centering
\begin{tabular}[h]{|p{2cm}|p{3.0cm}|p{3.0cm}|p{2.5cm}| }
\hline 
\multicolumn{4}{|c|}{$\mathrm{USp}(16)_{1}$ } \\
\hline
Line label & Highest weight & Quantum Dim. & Conf. Weight \\
\hline
0 & 0 & $d_{0} = 1$ & $h_{0} = 0$ \\
1 & $\mathbf{w}_{8}$ & $d_{1} = 1$ & $h_{1} = 2$ \\
2 & $\mathbf{w}_{7}$ & $d_{2} = \sqrt{2 + \phi}$ & $h_{2} = 77/40$ \\
3 & $\mathbf{w}_{1}$ & $d_{3} = \sqrt{2 + \phi}$ & $h_{3} = 17/40$ \\
4 & $\mathbf{w}_{6}$ & $d_{4} = (1 + \phi)$ & $h_{4} = 9/5$ \\
5 & $\mathbf{w}_{2}$ & $d_{5} = (1 + \phi)$ & $h_{5} = 4/5$ \\
6 & $\mathbf{w}_{5}$ & $d_{6} = \sqrt{3 + 4\phi}$ & $h_{6} = 13/8$ \\
7 & $\mathbf{w}_{3}$ & $d_{7} = \sqrt{3 + 4\phi}$ & $h_{7} = 9/8$ \\
8 & $\mathbf{w}_{4}$ & $d_{8} = 2\phi$ & $h_{8} = 7/5$ \\
\hline
\end{tabular}
\caption{Data of $\mathrm{USp}(16)_{1}$. In this table $\phi = (1 + \sqrt{5})/2$.}  \label{USp16lv1table}
\end{table}

\subsubsection*{$\mathrm{E}_{6,3} \times (\mathrm{G}_{2,1})^{3}$}

We provide now a final example involving abelian gauging of non-abelian MTCs. The example corresponds to entry number 32 in \cite{Schellekens:1992db} (see Appendix \ref{AppSchellekens}) and we find:
\begin{equation}
    \frac{\mathrm{E}_{6,3}}{\mathbb{Z}_{3}} \cong \mathrm{G}_{2,-1}^{3} - 48 \CSgrav.
\end{equation}
As before, we can check this expression using the spectra of $\mathrm{E}_{6,3}$ presented in Table \ref{E6lv3table} and applying the three-step procedure. Most lines are not gauge invariant, and so are removed from the spectrum. The exceptions are the lines in the representations $(3 \mathbf{w}_{1})$ and $(3 \mathbf{w}_{5})$ which are being gauged so are identified with the identity line, the lines $(\mathbf{w}_{6})$, $(\mathbf{w}_{1} + \mathbf{w}_{2})$ and $(\mathbf{w}_{4} + \mathbf{w}_{5})$ which identify with one another (they are in the same $\mathbb{Z}_{3}$ orbit), and the lines $(\mathbf{w}_{3})$ and $(\mathbf{w}_{1} + \mathbf{w}_{5})$ which are each a fixed point of the gauged $\mathbb{Z}_{3}$, and as such split into three lines. The result indeed matches the spectrum obtained from that of $\mathrm{G}_{2,1}$ presented in Appendix \ref{SummaryAnyonDataAppendix}.

\begin{table}[t]
\centering
\begin{tabular}[h]{|p{2cm}|p{3.0cm}|p{3.0cm}|p{2.5cm}| }
\hline 
\multicolumn{4}{|c|}{$\mathrm{E}_{6,3}$ } \\
\hline
Line label & Highest Weight & Quantum Dim. & Conf. Weight \\
\hline
0 & 0 & $d_{0} = 1$ & $h_{0} = 0$ \\
1 & 3$\mathbf{w}_{1}$ & $d_{1} = 1$ & $h_{1} = 2$ \\
2 & 3$\mathbf{w}_{5}$ & $d_{2} = 1$ & $h_{2} = 2$ \\
3 & $\mathbf{w}_{6}$ & $d_{3} = (1 + 2 \phi)$ & $h_{3} = 4/5$ \\
4 & $\mathbf{w}_{1} + \mathbf{w}_{2}$ & $d_{4} = (1 + 2 \phi)$ & $h_{4} = 9/5$ \\
5 & $\mathbf{w}_{4} + \mathbf{w}_{5}$ & $d_{5} = (1 + 2 \phi)$ & $h_{5} = 9/5$ \\
6 & $\mathbf{w}_{5}$ & $d_{6} = (2 + \phi)$ & $h_{6} = 26/45$ \\
7 & 2$\mathbf{w}_{1}$ & $d_{7} = (2 + \phi)$ & $h_{7} = 56/45$ \\
8 &  $\mathbf{w}_{1} + 2\mathbf{w}_{5}$ & $d_{8} = (2 + \phi)$ & $h_{8} = 86/45$ \\
9 &   $\mathbf{w}_{5} + \mathbf{w}_{6}$ & $d_{9} = (1 + 3 \phi)$ & $h_{9} = 13/9$ \\
10 &  $\mathbf{w}_{2}$ & $d_{10} = (1 + 3 \phi)$ & $h_{10} = 10/9$ \\
11 & $\mathbf{w}_{1} + \mathbf{w}_{4}$  & $d_{11} = (1 + 3 \phi)$ & $h_{11} = 16/9$ \\
12 & 2$\mathbf{w}_{5}$ & $d_{12} = (2 + \phi)$ & $h_{12} = 56/45$ \\
13 & $\mathbf{w}_{1}$ & $d_{13} = (2 + \phi)$ & $h_{13} = 26/45$ \\
14 &  $2\mathbf{w}_{1} + \mathbf{w}_{5}$ & $d_{14} = (2 + \phi)$ & $h_{14} = 86/45$ \\
15 & $\mathbf{w}_{4}$ & $d_{15} = (1 + 3 \phi)$ & $h_{15} = 10/9$ \\
16 & $\mathbf{w}_{1} + \mathbf{w}_{6}$ & $d_{16} = (1 + 3 \phi)$ & $h_{16} = 13/9$ \\
17 & $\mathbf{w}_{2} + \mathbf{w}_{5}$ & $d_{17} = (1 + 3 \phi)$ & $h_{17} = 16/9$ \\
18 & $\mathbf{w}_{3}$ & $d_{18} = 3 \phi$ & $h_{18} = 8/5$ \\
19 & $\mathbf{w}_{1} + \mathbf{w}_{5}$ & $d_{19} = 3(1 + \phi)$ & $h_{19} = 6/5$ \\
\hline
\end{tabular}
\caption{Data of $\mathrm{E}_{6,3}$. In this table $\phi = (1 + \sqrt{5})/2$.}  \label{E6lv3table}
\end{table}

\subsection{Non-Invertible Condensation Examples} \label{noninvertiblecondensationexamplessection}

\subsubsection*{$\mathrm{Spin}(18)_{2} \times \mathrm{SU}(8)_{1}$}

We move now to examples that require non-invertible (non-abelian) anyon condensation to establish Level-Rank dualities. The first example we consider is
\begin{equation} \label{firstnoninvertibleexample}
    \frac{\mathrm{Spin(18)_{2}}}{\mathcal{A}} = \mathrm{SU}(8)_{-1}
\end{equation}
which corresponds to entry number 50 in \cite{Schellekens:1992db} (see Appendix \ref{AppSchellekens}). The spectrum of $\mathrm{Spin(18)_{2}}$ is presented in Table \ref{Spin18lv2table}. Clearly, there are two bosons present in the spectrum, but condensing only the boson $(\mathbf{w}_{6})$ without multiplicity satisfies the quantum dimension constraint \eqref{quantumdimensionconstraint}. Alternatively, this is also the algebra obtained following the discussion in Section \ref{SectionDMNO}, and looking at the corresponding modular invariant obtained in \cite{Schellekens:1992db}. Thus, we expect the correct algebra to be
\begin{equation} \label{somealgebra}
    \mathcal{A} = (0) + (\mathbf{w}_{6}).
\end{equation}
Let us check this is indeed the case. First, we notice that the lines $(\mathbf{w}_{s} + \mathbf{w}_{c})$, $(\mathbf{w}_{1})$, $(\mathbf{w}_{2})$, $(\mathbf{w}_{7})$, $(\mathbf{w}_{4})$, $(\mathbf{w}_{5})$ are self-conjugate and do not split, which follows from the fact that their self-fusion contains a single copy of the identity line but does not contain $(\mathbf{w}_{6})$, e.g. $(\mathbf{w}_{s} + \mathbf{w}_{c}) \times (\mathbf{w}_{s} + \mathbf{w}_{c}) = (0) + (2\mathbf{w}_{1}) + (\mathbf{w}_{2})$. Thus, we have:
\begin{equation}
    (\mathbf{w}_{s} + \mathbf{w}_{c}) \rightarrow (\mathbf{w}_{s} + \mathbf{w}_{c})_{1}, \quad (\mathbf{w}_{1}) \rightarrow (\mathbf{w}_{1})_{1}, \quad (\mathbf{w}_{2}) \rightarrow (\mathbf{w}_{2})_{1}
\end{equation}
and similarly for $(\mathbf{w}_{7})$, $(\mathbf{w}_{4})$ and $(\mathbf{w}_{5})$.

\begin{table}[t]
\centering
\begin{tabular}[h]{|p{2cm}|p{3cm}|p{2.5cm}|p{2.5cm}|  }
\hline 
\multicolumn{4}{|c|}{$\mathrm{Spin}(18)_{2}$ } \\
\hline
Line label & Highest Weight & Quantum Dim. & Conf. Weight \\
\hline
0 & 0 & $d_{0} = 1$ & $h_{0} = 0$ \\
1 & 2$\mathbf{w}_{s}$ & $d_{1} = 1$ & $h_{1} = 9/4$ \\
2 & 2$\mathbf{w}_{1}$  & $d_{2} = 1$ & $h_{2} = 1$ \\
3 & 2$\mathbf{w}_{c}$  & $d_{3} = 1$ & $h_{3} = 9/4$ \\
4 & $\mathbf{w}_{c}$ & $d_{4} = 3$ & $h_{4} = 17/16$ \\
5 & $\mathbf{w}_{s}$ & $d_{5} = 3$ & $h_{5} = 17/16$ \\
6 & $\mathbf{w}_{1} + \mathbf{w}_{s}$ & $d_{6} = 3$ & $h_{6} = 25/16$ \\
7 & $\mathbf{w}_{1} + \mathbf{w}_{c}$ & $d_{7} = 3$ & $h_{7} = 25/16$ \\
8 & $\mathbf{w}_{s} + \mathbf{w}_{c}$ & $d_{8} = 2$ & $h_{8} = 20/9$ \\
9 & $\mathbf{w}_{1}$ & $d_{9} = 2$ & $h_{9} = 17/36$ \\
10 & $\mathbf{w}_{2}$ & $d_{10} = 2$ & $h_{10} = 8/9$ \\
11 & $\mathbf{w}_{7}$ & $d_{11} = 2$ & $h_{11} = 77/36$ \\
12 & $\mathbf{w}_{6}$ & $d_{12} = 2$ & $h_{12} = 2$ \\
13 & $\mathbf{w}_{3}$ & $d_{13} = 2$ & $h_{13} = 5/4$ \\
14 & $\mathbf{w}_{4}$ & $d_{14} = 2$ & $h_{14} = 14/9$ \\
15 & $\mathbf{w}_{5}$ & $d_{15} = 2$ & $h_{15} = 65/36$ \\
\hline
\end{tabular}
\caption{Data of $\mathrm{Spin}(18)_{2}$.}  \label{Spin18lv2table}
\end{table}

Now, if we analyze $(\mathbf{w}_{s} + \mathbf{w}_{c}) \times (\mathbf{w}_{2}) = (\mathbf{w}_{s} + \mathbf{w}_{c}) + (\mathbf{w}_{6}) \rightarrow (0) + \ldots$, we deduce that $(\mathbf{w}_{s} + \mathbf{w}_{c})_{1}$ and $(\mathbf{w}_{2})_{1}$ must be identified with each other since $(\mathbf{w}_{6})$ condenses. Following similar steps, we deduce the identifications:
\begin{equation}
    (\mathbf{w}_{s} + \mathbf{w}_{c})_{1} \cong (\mathbf{w}_{2})_{1} \cong (\mathbf{w}_{4})_{1}, \quad (\mathbf{w}_{1})_{1} \cong (\mathbf{w}_{7})_{1} \cong (\mathbf{w}_{5})_{1}.
\end{equation}
Meanwhile, we see that $(\mathbf{w}_{6}) \times (\mathbf{w}_{6}) = (0) + (2\mathbf{w}_{1}) + (\mathbf{w}_{6}) \rightarrow 2 \, (0) + \ldots$ so $(\mathbf{w}_{6})$ splits into two components (one of which is the identity line since $(\mathbf{w}_{6})$ condenses), and noticing that $(2 \mathbf{w}_{1}) \times (\mathbf{w}_{6}) = (\mathbf{w}_{6}) \rightarrow (0) + \ldots$, we deduce that $(2\mathbf{w}_{1})_{1}$ belongs in the splitting of $(\mathbf{w}_{6})$. Thus, we find $(\mathbf{w}_{6}) \rightarrow (0) + (2\mathbf{w}_{1})_{1}$, and a similar argument shows that $(\mathbf{w}_{3}) \rightarrow (2\mathbf{w}_{s})_{1} + (2\mathbf{w}_{c})_{1}$. Similarly, we deduce that $(\mathbf{w}_{c}), (\mathbf{w}_{s}), (\mathbf{w}_{1} + \mathbf{w}_{s}), (\mathbf{w}_{1} + \mathbf{w}_{c})$ splits into two components, but since now the lines are not self-conjugate, we find
\begin{align}
    (\mathbf{w}_{c}) \rightarrow (\mathbf{w}_{c})_{1} + (\mathbf{w}_{c})_{2}, \quad &(\mathbf{w}_{s}) \rightarrow \overline{(\mathbf{w}_{c})}_{1} + \overline{(\mathbf{w}_{c})}_{2}, \\[0.2cm] (\mathbf{w}_{1} + \mathbf{w}_{s}) \rightarrow (\mathbf{w}_{1} + \mathbf{w}_{s})_{1} + (\mathbf{w}_{1} + \mathbf{w}_{s})_{2}, \quad &(\mathbf{w}_{1} + \mathbf{w}_{c}) \rightarrow \overline{(\mathbf{w}_{1} + \mathbf{w}_{s})}_{1} + \overline{(\mathbf{w}_{1} + \mathbf{w}_{s})}_{2} 
\end{align}
Next, we need to decide how to assign quantum dimensions on the right hand sides. To do this we calculate $(\mathbf{w}_{1} + \mathbf{w}_{s}) \times (\mathbf{w}_{1} + \mathbf{w}_{c}) = (0) + (\mathbf{w}_{s} + \mathbf{w}_{c}) + (\mathbf{w}_{2}) + (\mathbf{w}_{4}) + (\mathbf{w}_{6})$ so that we can pair up conjugates together. We obtain, after splitting:
\begin{align}
    (\mathbf{w}_{1} + \mathbf{w}_{s})_{1} &\times \overline{(\mathbf{w}_{1} + \mathbf{w}_{s})}_{1} + (\mathbf{w}_{1} + \mathbf{w}_{s})_{1} \times \overline{(\mathbf{w}_{1} + \mathbf{w}_{s})}_{2} + (\mathbf{w}_{1} + \mathbf{w}_{s})_{2} \times \overline{(\mathbf{w}_{1} + \mathbf{w}_{s})}_{1} \nonumber \\[0.3cm] & + (\mathbf{w}_{1} + \mathbf{w}_{s})_{2} \times \overline{(\mathbf{w}_{1} + \mathbf{w}_{s})}_{2} = 2 \, (0) + (2\mathbf{w}_{1})_{1} + 3 \, (\mathbf{w}_{s} + \mathbf{w}_{c})_{1}
\end{align}
The identity lines on the right hand side must obviously be assigned to $(\mathbf{w}_{1} + \mathbf{w}_{s})_{1} \times \overline{(\mathbf{w}_{1} + \mathbf{w}_{s})}_{1}$ and $(\mathbf{w}_{1} + \mathbf{w}_{s})_{2} \times \overline{(\mathbf{w}_{1} + \mathbf{w}_{s})}_{2}$. Now, since the quantum dimension of $(\mathbf{w}_{1} + \mathbf{w}_{s})_{1} \times \overline{(\mathbf{w}_{1} + \mathbf{w}_{s})}_{2}$ and $(\mathbf{w}_{1} + \mathbf{w}_{s})_{2} \times \overline{(\mathbf{w}_{1} + \mathbf{w}_{s})}_{1}$ must agree the only consistent possibility is that $(\mathbf{w}_{1} + \mathbf{w}_{s})_{1} \times \overline{(\mathbf{w}_{1} + \mathbf{w}_{s})}_{2} = (\mathbf{w}_{1} + \mathbf{w}_{s})_{2} \times \overline{(\mathbf{w}_{1} + \mathbf{w}_{s})}_{1} = (\mathbf{w}_{s} + \mathbf{w}_{c})_{1}$. Now, we could try to assign $(\mathbf{w}_{1} + \mathbf{w}_{s})_{1} \times \overline{(\mathbf{w}_{1} + \mathbf{w}_{s})}_{1} = (0) + (2\mathbf{w}_{1})_{1}$ and $(\mathbf{w}_{1} + \mathbf{w}_{s})_{2} \times \overline{(\mathbf{w}_{1} + \mathbf{w}_{s})}_{2} = (0) + (\mathbf{w}_{s} + \mathbf{w}_{c})_{1}$, but this is inconsistent with the conservation of quantum dimension $d_{(\mathbf{w}_{1} + \mathbf{w}_{s})} = d_{(\mathbf{w}_{1} + \mathbf{w}_{s})_{1}} + d_{(\mathbf{w}_{1} + \mathbf{w}_{s})_{2}}$. Then, the only consistent fusion rules are:
\begin{align}
    &(\mathbf{w}_{1} + \mathbf{w}_{s})_{1} \times \overline{(\mathbf{w}_{1} + \mathbf{w}_{s})}_{1} = (0), \quad (\mathbf{w}_{1} + \mathbf{w}_{s})_{2} \times \overline{(\mathbf{w}_{1} + \mathbf{w}_{s})}_{2} = (0) + (2\mathbf{w}_{1})_{1} + (\mathbf{w}_{s} + \mathbf{w}_{c})_{1}, \nonumber \\[0.3cm] &(\mathbf{w}_{1} + \mathbf{w}_{s})_{1} \times \overline{(\mathbf{w}_{1} + \mathbf{w}_{s})}_{2} = (\mathbf{w}_{1} + \mathbf{w}_{s})_{2} \times \overline{(\mathbf{w}_{1} + \mathbf{w}_{s})}_{1} = (\mathbf{w}_{s} + \mathbf{w}_{c})_{1},
\end{align}
which in particular implies that $d_{(\mathbf{w}_{1} + \mathbf{w}_{s})_{1}} = 1$  and $d_{(\mathbf{w}_{1} + \mathbf{w}_{s})_{2}} = 2$. A similar analysis yields $d_{(\mathbf{w}_{c})_{1}} = 1$  and $d_{(\mathbf{w}_{c})_{2}} = 2$. It is also easy to check that no other lines identify with any of $(\mathbf{w}_{c})_{1}, (\mathbf{w}_{c})_{2}, (\mathbf{w}_{1} + \mathbf{w}_{s})_{1}, (\mathbf{w}_{1} + \mathbf{w}_{s})_{2}$. There could be identification between lines in this set, however.

To check this, consider the fusion $(\mathbf{w}_{s}) \times (\mathbf{w}_{1} + \mathbf{w}_{s}) = (2\mathbf{w}_{1}) + (\mathbf{w}_{s} + \mathbf{w}_{c}) + (\mathbf{w}_{2}) + (\mathbf{w}_{4}) + (\mathbf{w}_{6})$. After splitting:
\begin{align}
    \overline{(\mathbf{w}_{c})}_{1} \times (\mathbf{w}_{1} + \mathbf{w}_{s})_{1} &+ \overline{(\mathbf{w}_{c})}_{1} \times (\mathbf{w}_{1} + \mathbf{w}_{s})_{2} + \overline{(\mathbf{w}_{c})}_{2} \times (\mathbf{w}_{1} + \mathbf{w}_{s})_{1} \nonumber \\[0.3cm] &+ \overline{(\mathbf{w}_{c})}_{2} \times (\mathbf{w}_{1} + \mathbf{w}_{s})_{2} = 0 + 2 \, (2\mathbf{w}_{1})_{1} + 3 \, (\mathbf{w}_{s} + \mathbf{w}_{c})_{1}.
\end{align}
Thus, we must have either $ \overline{(\mathbf{w}_{c})}_{1} \times (\mathbf{w}_{1} + \mathbf{w}_{s})_{1} = (0)$ or $ \overline{(\mathbf{w}_{c})}_{2} \times (\mathbf{w}_{1} + \mathbf{w}_{s})_{2} = (0) + \cdots$. We can rule out the first possibility as it would imply we have to identify $(\mathbf{w}_{1} + \mathbf{w}_{s})_{1} \cong (\mathbf{w}_{c})_{1}$. However, $(2\mathbf{w}_{1}) \times (\mathbf{w}_{c}) = (\mathbf{w}_{1} + \mathbf{w}_{s}) \Rightarrow (2\mathbf{w}_{1})_{1} \times (\mathbf{w}_{c})_{1} + (2\mathbf{w}_{1})_{1} \times (\mathbf{w}_{c})_{2} = (\mathbf{w}_{1} + \mathbf{w}_{s})_{1} + (\mathbf{w}_{1} + \mathbf{w}_{s})_{2} = (\mathbf{w}_{c})_{1} + (\mathbf{w}_{1} + \mathbf{w}_{s})_{2}$. Then, matching quantum dimensions $(2\mathbf{w}_{1})_{1} \times (\mathbf{w}_{c})_{1} = (\mathbf{w}_{c})_{1} \Rightarrow (2\mathbf{w}_{1})_{1} = (0)$, which is incorrect since $(2\mathbf{w}_{1})$ is not condensing. We deduce then that $\overline{(\mathbf{w}_{c})}_{2} \times (\mathbf{w}_{1} + \mathbf{w}_{s})_{2} = (0) + \cdots$ which implies the identification $(\mathbf{w}_{c})_{2} \cong (\mathbf{w}_{1} + \mathbf{w}_{s})_{2}$. Organizing quantum dimensions, it is straightforward to see the resulting fusion rules are
\begin{align}
    &\overline{(\mathbf{w}_{c})}_{2} \times (\mathbf{w}_{1} + \mathbf{w}_{s})_{2} = (0) + (2\mathbf{w}_{1})_{1} + (\mathbf{w}_{s} + \mathbf{w}_{c})_{1}, \quad \overline{(\mathbf{w}_{c})}_{1} \times (\mathbf{w}_{1} + \mathbf{w}_{s})_{1} = (2\mathbf{w}_{1})_{1}, \nonumber \\[0.3cm] &\overline{(\mathbf{w}_{c})}_{1} \times (\mathbf{w}_{1} + \mathbf{w}_{s})_{2} = \overline{(\mathbf{w}_{c})}_{2} \times (\mathbf{w}_{1} + \mathbf{w}_{s})_{1} = (\mathbf{w}_{s} + \mathbf{w}_{c})_{1},
\end{align}
and we identify $(\mathbf{w}_{c})_{2} \cong (\mathbf{w}_{1} + \mathbf{w}_{s})_{2}$. All in all, we obtain the splitting and identification summarized in Table \ref{Spin18splittings}.

\begin{table}[t]
\centering
\begin{tabular}[h]{|p{2.5cm}|p{2.0cm}|p{4.0cm}| }
\hline 
\multicolumn{3}{|c|}{$\textrm{Spin}(18)_2$ } \\
\hline
$\textrm{Spin}(18)_2$ line  & $h \mod 1$ & Splitting\\
\hline
$(0)$ &  $0$ & $(0)$ \\
 $(2\mathbf{w}_s)$ & $1/4$ & $(2 \mathbf{w}_s)_1  $ \\
$(2\mathbf{w}_1)$ & $0$ & $ (2 \mathbf{w}_1)_1 $ \\
$(2\mathbf{w}_c)$ & $1/4$ & $(2 \mathbf{w}_c)_1   $ \\
$(\mathbf{w}_c)$ & $1/16$ & $(\mathbf{w}_c)_1+(\mathbf{w}_c)_2 $ \\
$(\mathbf{w}_s)$ & $1/16$ & $ \overline{(\mathbf{w}_c)_1}+\overline{(\mathbf{w}_c)_2} $ \\
$(\mathbf{w}_1+\mathbf{w}_s)$ & $9/16$ & $ (\mathbf{w}_1+\mathbf{w}_s)_1+(\mathbf{w}_c)_2$ \\
$(\mathbf{w}_1+\mathbf{w}_c)$ & $9/16$ & $ \overline{(\mathbf{w}_1+\mathbf{w}_s)_1}+\overline{(\mathbf{w}_c)_2} $ \\
$(\mathbf{w}_s+\mathbf{w}_c)$ & $2/9$ & $(\mathbf{w}_s+\mathbf{w}_c)_1 $ \\
$(\mathbf{w}_1)$ & $17/36$ & $ (\mathbf{w}_1)_1 $ \\
$(\mathbf{w}_2)$ & $8/9$ & $(\mathbf{w}_s+\mathbf{w}_c)_1 $ \\
$(\mathbf{w}_7)$ & $5/36$ & $(\mathbf{w}_1)_1$  \\
$(\mathbf{w}_6)$ &  $0$ &  $  (0)+(2 \mathbf{w}_1)_1$ \\
$(\mathbf{w}_3)$ & $1/4$ & $ (2 \mathbf{w}_s)_1+(2 \mathbf{w}_c)_1$ \\
$(\mathbf{w}_4)$ & $5/9$ &$ (\mathbf{w}_s+\mathbf{w}_c)_1$ \\
$(\mathbf{w}_5)$ & $29/36$ &  $ (\mathbf{w}_1)_1$ \\
\hline
\end{tabular}
\caption{Splitting of $\textrm{Spin}(18)_2$ lines after condensation of ${\cal A}=(0)+(\mathbf{w}_6)$ (after identifications are implemented).} \label{Spin18splittings}
\end{table}

Clearly, all lines that are not abelian confine, and the abelian ones have the correct spin to reproduce $\mathrm{SU}(8)_{-1}$. All that remains is to verify the fusion ring. To do this we consider the fusion $(\mathbf{w}_{c}) \times (\mathbf{w}_{c}) = (2\mathbf{w}_{c}) + (\mathbf{w}_{1}) + (\mathbf{w}_{3}) + (\mathbf{w}_{5}) + (\mathbf{w}_{7})$, and consider the equation resulting after splitting:
\begin{align}
    (\mathbf{w}_{c})_{1} \times (\mathbf{w}_{c})_{1} &+ (\mathbf{w}_{c})_{1} \times (\mathbf{w}_{c})_{2} + (\mathbf{w}_{c})_{2} \times (\mathbf{w}_{c})_{1} \nonumber \\[0.3cm] + & (\mathbf{w}_{c})_{2} \times (\mathbf{w}_{c})_{2} = (2\mathbf{w}_{s})_{1} + 2 \, (2\mathbf{w}_{c})_{1} + 3 \, (\mathbf{w}_{1})_{1}.
\end{align}
Matching quantum dimensions we either have $(\mathbf{w}_{c})_{1} \times (\mathbf{w}_{c})_{1} = (2\mathbf{w}_{s})_{1}$ or $(\mathbf{w}_{c})_{1} \times (\mathbf{w}_{c})_{1} = (2\mathbf{w}_{c})_{1}$. The first option is inconsistent, since otherwise we have $(\mathbf{w}_{c})_{1} \times ((\mathbf{w}_{c})_{1} \times (2\mathbf{w}_{c})) = (\mathbf{w}_{c})_{1} \times \overline{(\mathbf{w}_{1} + \mathbf{w}_{s})}_{1} = (0)$ and thus we would have to identify $(\mathbf{w}_{c})_{1}$ with $(\mathbf{w}_{1} + \mathbf{w}_{s})_{1}$ which we have already deduced does not happen. Thus, we have $(\mathbf{w}_{c})_{1} \times (\mathbf{w}_{c})_{1} = (2 \mathbf{w}_{c})_{1}$, from which is easy to deduce the rest of the fusion ring, and it indeed reproduces that of $\mathrm{SU}(8)_{-1}$. This finishes the check that $\mathrm{Spin}(18)_{2}/\mathcal{A} \cong \mathrm{SU}(8)_{-1}$ where the algebra $\mathcal{A}$ is defined in \eqref{somealgebra}.

We can generalize the preceding example in the following manner. First, in order to guess a general pattern, it is instructive to notice that
\begin{equation} \label{firstequality}
    \mathrm{Spin}(2)_{2} = \mathrm{U}(1)_{8} \cong \mathrm{SU}(8)_{-1},
\end{equation}
where the first equality comes from the fact that $\mathrm{Spin(2)}_{k} = \mathrm{U}(1)_{4k}$ for general integer $k$, and the second expression follows from the bosonic Level-Rank duality $\mathrm{SU}(N)_{k} \cong \mathrm{U}(k)_{-N}$ which holds as long as $N$ is even and $N k = 0 \mod 8$ \cite{Aharony:2016jvv, Hsin:2019gvb}.\footnote{As spin TQFTs the Level-Rank duality holds for arbitrary positive integer $N$ and $k$ \cite{Hsin:2016blu}.} In particular, notice that both \eqref{firstequality} and \eqref{firstnoninvertibleexample} fit in the following infinite pattern:
\begin{equation} \label{infinitepattern}
    \frac{\mathrm{Spin}(2n^{2})_{2}}{\mathcal{A}_{n}} \cong \mathrm{SU}(8)_{-1},
\end{equation}
for odd $n$. For $n=1$ the algebra on the left-hand side is trivial and the expression reproduces \eqref{firstequality}, while for $n=3$ it reproduces \eqref{firstnoninvertibleexample} with the algebra given by \eqref{somealgebra}.

\begin{table}[t]
\centering
\begin{tabular}[h]{|p{2cm}|p{3cm}|p{2.5cm}|p{3.1cm}|  }
\hline 
\multicolumn{4}{|c|}{$\mathrm{Spin}(2m)_{2}$ } \\
\hline
Line label & Highest Weight & Quantum Dim. & Conformal Weight \\
\hline
0 & 0 & $d_{0} = 1$ & $h_{0} = 0$ \\
1 & 2$\mathbf{w}_{s}$ & $d_{1} = 1$ & $h_{1} = m/4$ \\
2 & 2$\mathbf{w}_{1}$  & $d_{2} = 1$ & $h_{2} = 1$ \\
3 & 2$\mathbf{w}_{c}$  & $d_{3} = 1$ & $h_{3} = m/4$ \\
4 & $\mathbf{w}_{s}$ & $d_{4} = \sqrt{m}$ & $h_{4} = \frac{(2m-1)}{16}$ \\
5 & $\mathbf{w}_{c}$ & $d_{5} = \sqrt{m}$ & $h_{5} = \frac{(2m-1)}{16}$ \\
6 & $\mathbf{w}_{1} + \mathbf{w}_{s}$ & $d_{6} = \sqrt{m}$ & $h_{6} = \frac{(2m-1)}{16} + \frac{1}{2}$ \\
7 & $\mathbf{w}_{1} + \mathbf{w}_{c}$ & $d_{7} = \sqrt{m}$ & $h_{7} = \frac{(2m-1)}{16} + \frac{1}{2}$ \\
8 & $\mathbf{w}_{s} + \mathbf{w}_{c}$ & $d_{8} = 2$ & $h_{8} = \frac{m^{2}-1}{4m}$ \\
8 + $i$ & $\mathbf{w}_{i}$ & $d_{8+i} = 2$ & $h_{8 + i} = \frac{i}{2} - \frac{i^{2}}{4m}$ \\
\hline
\end{tabular}
\caption{Data of $\mathrm{Spin}(2m)_{2}$.}  \label{Spin2mlv2table}
\end{table}

To check that the previous pattern is correct, consider the following proposal for the algebra:
\begin{equation} \label{algebraAn}
    \mathcal{A}_{n} = (0) + \sum_{i=1}^{(n-1)/2}(\mathbf{w}_{2ni}).
\end{equation}
Indeed, it is straightforward to verify using the MTC data presented in Table \ref{Spin2mlv2table} that all the lines in $\mathrm{Spin}(2n^{2})_{2}$ appearing above are bosons. Moreover, it is also easy to verify from the table that the quantum dimension constraint \eqref{quantumdimensionconstraint} is satisfied for the collection of bosons above.

As a further check, we recall that $\mathrm{SU}(8)_{1}$ has a gaugable $\mathbb{Z}_{2}$ one-form symmetry that leads to the duality \cite{Cordova:2018qvg}
\begin{equation} \label{knownlevelrankduality}
     \frac{\mathrm{SU}(8)_{-1}}{\mathbb{Z}_{2}} \cong \mathrm{SU}(2)_{1}.
\end{equation}
If \eqref{infinitepattern} is correct, we deduce from this expression that a topological interface must exist between $\mathrm{Spin}(2n^{2})_{2}$ and $\mathrm{SU}(2)_{1}$ for any odd $n$. Indeed, we can directly prove the existence of this topological interface by alternatively noticing that the abelian TQFT
\begin{equation}
    \mathrm{SU}(2n^{2})_{1} \times \mathrm{SU}(2)_{-1}
\end{equation}
has a Lagrangian subgroup generated by the line of charges $(n,1)$. Gauging charge conjugation in $\mathrm{SU}(2n^{2})_{1}$ gives $\mathrm{Spin}(2n^{2})_{2}$, and thus upon unfolding we find the desired topological interface between $\mathrm{Spin}(2n^{2})_{2}$ and $\mathrm{SU}(2)_{1}$. Using Proposition 1, we also find then that there must exist an algebra $\mathcal{B}_{n}$ such that the  duality \eqref{knownlevelrankduality} can further be extended to
\begin{equation} \label{extendedlevelrankspintosu2}
    \frac{\mathrm{Spin}(2n^{2})_{2}}{\mathcal{B}_{n}} \cong \frac{\mathrm{SU}(8)_{-1}}{\mathbb{Z}_{2}} \cong \mathrm{SU}(2)_{1}.
\end{equation}

The existence of the algebra $\mathcal{A}_{n}$ implies as well a corresponding algebra $\mathcal{B}_{n}$ taking $\mathrm{Spin}(2n^{2})_{2}$ to $\mathrm{SU}(2)_{1}$. To see this, we merely stack the interface generated by $\mathcal{A}_{n}$ and that generated by the $\mathbb{Z}_{2}$ gauging of $\mathrm{SU}(8)_{-1}$ to $\mathrm{SU}(2)_{1}$. Following the discussion in Section \ref{SectionDMNO}, we recover the following form for the algebra $\mathcal{B}_{n}$:
\begin{equation}
    \mathcal{B}_{n} = (0) + (2\mathbf{w}_{1}) + 2 \sum_{i=1}^{(n-1)/2}(\mathbf{w}_{2ni}).
\end{equation}
As above, we can verify that all the lines in the decomposition are bosons and that the quantum dimension constraint \eqref{quantumdimensionconstraint} is obeyed. That we can obtain such an algebra is unsurprising, since it is guaranteed by the argument above \eqref{extendedlevelrankspintosu2}. The important observation is that the existence of such an algebra can also be deduced independently from the existence of the algebra $\mathcal{A}_{n}$, which provides a non-trivial verification of \eqref{infinitepattern}.

\subsubsection*{$\mathrm{E}_{7,3} \times \mathrm{SU}(6)_{1}$}

Finally, we end our series of detailed examples with a last case where we need to perform non-abelian anyon condensation. This last example is taken from entry number 45 of \cite{Schellekens:1992db} (see Appendix \ref{AppSchellekens}), and consists of the duality
\begin{equation}
    \frac{\mathrm{E}_{7,3}}{\mathcal{A}} \cong \mathrm{SU}(6)_{-1}.
\end{equation}
The data of $\mathrm{E}_{7,3}$ is presented in Table \ref{E7lv3table}. Following our discussion, we condense the line $(\mathbf{w}_{6} + \mathbf{w}_{7})$ in the notation of Table \ref{E7lv3table}, which is the only non-trivial boson in the spectrum. Moreover, we condense it without multiplicity as this is the only possibility consistent with the quantum dimension constraint \eqref{quantumdimensionconstraint}. Alternatively, this is also the algebra obtained following the discussion in Section \ref{SectionDMNO}, and looking at the corresponding modular invariant obtained in \cite{Schellekens:1992db}. Thus, we have:
\begin{equation}
    \mathcal{A} = (0) + (\mathbf{w}_{6} + \mathbf{w}_{7}).
\end{equation}
Let us proceed to show this gives indeed $\mathrm{SU}(6)_{-1}$. First, we notice that all lines of $\mathrm{E}_{7,3}$ are self-conjugate, so we can check if they split or not calculating fusions of the form $i \times i$. We find the splittings (before any identifications):
\begin{table}[t]
\centering
\begin{tabular}[h]{|p{2cm}|p{3cm}|p{2.5cm}|p{2.5cm}| }
\hline 
\multicolumn{4}{|c|}{$\mathrm{E}_{7,3}$ } \\
\hline
Line label & Highest weight & Quantum Dim. & Conf. Weight \\
\hline
0 & 0  & $d_{0} = 1$ & $h_{0} = 0$ \\
1 & 3$\mathbf{w}_{6}$ & $d_{1} = 1$ & $h_{1} = 9/4$ \\
2 & $\mathbf{w}_{6} + \mathbf{w}_{7}$  & $d_{2} = d$ & $h_{2} = 2$ \\
3 & $\mathbf{w}_{7}$ & $d_{3} = d$ & $h_{3} = 5/4$ \\
4 & 2$\mathbf{w}_{6}$ & $d_{4} = d-1$ & $h_{4} = 10/7$ \\
5 & $\mathbf{w}_{6}$ & $d_{5} = d-1$ & $h_{5} = 19/28$ \\
6 & $\mathbf{w}_{5}$ & $d_{6} = d+1$ & $h_{6} = 4/3$ \\
7 & $\mathbf{w}_{1} + \mathbf{w}_{6}$ & $d_{7} = d+1$ & $h_{7} = 19/12$ \\
8 & $\mathbf{w}_{1}$ & $d_{8} = d-1$ & $h_{8} = 6/7$ \\
9 & $\mathbf{w}_{5} + \mathbf{w}_{6}$ & $d_{9} = d-1$ & $h_{9} = 59/28$ \\
10 & $\mathbf{w}_{2}$ & $d_{10} = d-1$ & $h_{10} = 12/7$ \\
11 & $\mathbf{w}_{4}$ & $d_{11} = d-1$ & $h_{11} = 55/28$ \\
\hline
\end{tabular} 
\caption{Data of $\mathrm{E}_{7,3}$, where $d = \frac{1}{2}(5 + \sqrt{21})$.}  \label{E7lv3table}
\end{table}
\begin{align}
    & (0) \rightarrow (0), \quad (3\mathbf{w}_{6}) \rightarrow (3\mathbf{w}_{6})_{1}, \quad (\mathbf{w}_{6} + \mathbf{w}_{7}) \rightarrow (0) + (\mathbf{w}_{6} + \mathbf{w}_{7})_{2}, \quad (\mathbf{w}_{7}) \rightarrow (\mathbf{w}_{7})_{1} + (\mathbf{w}_{7})_{2}, \nonumber \\[0.3cm] &(2\mathbf{w}_{6}) \rightarrow (2\mathbf{w}_{6})_{1}, \quad (\mathbf{w}_{6}) \rightarrow (\mathbf{w}_{6})_{1}, \quad (\mathbf{w}_{5}) \rightarrow (\mathbf{w}_{5})_{1} + (\mathbf{w}_{5})_{2} + (\mathbf{w}_{5})_{3} \nonumber \\[0.35cm] & (\mathbf{w}_{1} + \mathbf{w}_{6}) \rightarrow (\mathbf{w}_{1} + \mathbf{w}_{6})_{1} + (\mathbf{w}_{1} + \mathbf{w}_{6})_{2} + (\mathbf{w}_{1} + \mathbf{w}_{6})_{3}, \quad (\mathbf{w}_{1}) \rightarrow (\mathbf{w}_{1})_{1}, \nonumber \\[0.3cm] &(\mathbf{w}_{5} + \mathbf{w}_{6}) \rightarrow (\mathbf{w}_{5} + \mathbf{w}_{6})_{1}, \quad (\mathbf{w}_{2}) \rightarrow (\mathbf{w}_{2})_{1}, \quad (\mathbf{w}_{4}) \rightarrow (\mathbf{w}_{4})_{1}.
\end{align}
It is easier to identify the lines that do not split. For instance, if we compute
\begin{equation}
   (2\mathbf{w}_{6}) \times (\mathbf{w}_{1}) = (\mathbf{w}_{6} + \mathbf{w}_{7}) + (2\mathbf{w}_{6}) + (\mathbf{w}_{5}) \rightarrow (0) + \cdots,
\end{equation}
we find $(2\mathbf{w}_{6})_{1}$ must be conjugate to $(\mathbf{w}_{1})_{1}$, but since the lines do not split and are self-conjugate before condensation, we find that necessarily
\begin{equation}
    (2\mathbf{w}_{6})_{1} \cong (\mathbf{w}_{1})_{1}.
\end{equation}
Proceeding in the same way:
\begin{equation}
    (2\mathbf{w}_{6})_{1} \cong (\mathbf{w}_{1})_{1} \cong (\mathbf{w}_{2})_{1}, \quad (\mathbf{w}_{6})_{1} \cong (\mathbf{w}_{5} + \mathbf{w}_{6})_{1} \cong (\mathbf{w}_{4})_{1},
\end{equation}
and the lines clearly confine.

We now check if they lines that do not split appear as a component of the lines that do split. For example:
\begin{equation}
    (3\mathbf{w}_{6}) \times (\mathbf{w}_{7}) = (\mathbf{w}_{6} + \mathbf{w}_{7}) \rightarrow (0) + \cdots,
\end{equation}
Thus, one of the components of $(\mathbf{w}_{7})$ must be identified with $(3\mathbf{w}_{6})_{1}$ since the latter line is self-conjugate: $(\mathbf{w}_{7})_{1} \cong (3\mathbf{w}_{6})_{1}$. Proceeding similarly with the rest of the lines, we obtain the pattern summarized in Table \ref{E7lv3splittings}.

We also notice that $(\mathbf{w}_{5}) \times (\mathbf{w}_{1} + \mathbf{w}_{6})$ in $\mathrm{E}_{7,3}$ does not contain either (0) or $(\mathbf{w}_{6} + \mathbf{w}_{7})$, and therefore $(\mathbf{w}_{5})_{2}, (\mathbf{w}_{5})_{3},(\mathbf{w}_{1} + \mathbf{w}_{6})_{2}, (\mathbf{w}_{1} + \mathbf{w}_{6})_{3}$ are not identified with other lines or between each other. Checking quantum dimensions, we see that $d_{(\mathbf{w}_{5})_{2}} = d_{(\mathbf{w}_{5})_{3}} = d_{(\mathbf{w}_{1} + \mathbf{w}_{6})_{2}} = d_{(\mathbf{w}_{1} + \mathbf{w}_{6})_{3}} = 1$. Moreover, it is easy to see that none of $ (0), (3 \mathbf{w}_{6})_{1}, (\mathbf{w}_{5})_{2}, (\mathbf{w}_{5})_{3},(\mathbf{w}_{1} + \mathbf{w}_{6})_{2},(\mathbf{w}_{1} + \mathbf{w}_{6})_{3}$ confine, and they have exactly the correct topological spins for them to be recognized as the lines in $\mathrm{SU}(6)_{-1}$. All that remains is to check the fusion ring.

\begin{table}[t]
\centering
\begin{tabular}[h]{|p{2.0cm}|p{2cm}|p{5.5cm}| }
\hline 
\multicolumn{3}{|c|}{$\mathrm{E}_{7,3}$ } \\
\hline
$\mathrm{E}_{7,3}$ line & $h$ mod $1$ & Splitting\\
\hline
$(0)$ & $0$ & $(0)$ \\
$(3 \mathbf{w}_6)$ & $1/4$ & $(3 \mathbf{w}_6)_1$ \\
$(\mathbf{w}_6+\mathbf{w}_7)$ & $0$ & $(0)+(2 \mathbf{w}_6)_1$ \\
$(\mathbf{w}_7)$ & $1/4$ & $(3 \mathbf{w}_6)_1+(\mathbf{w}_6)_1$ \\
$(2 \mathbf{w}_6)$ & $3/7$ & $(2 \mathbf{w}_6)_1$ \\
$(\mathbf{w}_6)$ & $19/28$ & $(\mathbf{w}_6)_1$ \\
$(\mathbf{w}_5)$ & $1/3$ & $(2 \mathbf{w}_6)_1 + (\mathbf{w}_5)_2+(\mathbf{w}_5)_3$ \\
$(\mathbf{w}_1+\mathbf{w}_6)$ & $7/12$ & $(\mathbf{w}_6)_1+(\mathbf{w}_1+\mathbf{w}_6)_2+(\mathbf{w}_1+\mathbf{w}_6)_3$ \\
$(\mathbf{w}_1)$ & $6/7$ & $( 2 \mathbf{w}_6)_1$ \\
$(\mathbf{w}_5+\mathbf{w}_6)$ & $3/28$ & $(\mathbf{w}_6)_1$ \\
$(\mathbf{w}_2)$ & $5/7$ & $(2 \mathbf{w}_6)_1$ \\
$(\mathbf{w}_4)$ & $27/28$ & $(\mathbf{w}_6)_1$ \\
\hline
\end{tabular}
\caption{Splitting of $\mathrm{E}_{7,3}$ lines after condensation of ${\cal A}=(0)+(\mathbf{w}_6+\mathbf{w}_7)$ (after identifications are implemented).} \label{E7lv3splittings}
\end{table}

To do this, we first need to compute $(2\mathbf{w}_{6})_{1} \times (\mathbf{w}_{5})_{2} = (\mathbf{w}_{5})_{2} \times (2\mathbf{w}_{6})_{1}$ and $(2\mathbf{w}_{6})_{1} \times (\mathbf{w}_{5})_{3} = (\mathbf{w}_{5})_{3} \times (2\mathbf{w}_{6})_{1}$, which we can obtain by computing $ (2\mathbf{w}_{6}) \times (\mathbf{w}_{5})$ and comparing before and after splitting. It is straightforward to check that this gives:
\begin{equation}
    (2\mathbf{w}_{6})_{1} \times (\mathbf{w}_{5})_{2} = (2\mathbf{w}_{6})_{1} \times (\mathbf{w}_{5})_{3} = (\mathbf{w}_{5})_{2} \times (2\mathbf{w}_{6})_{1} = (\mathbf{w}_{5})_{3} \times (2\mathbf{w}_{6})_{1} = (2\mathbf{w}_{6})_{1}.
\end{equation}
Now, to find the fusion ring involving $(\mathbf{w}_{5})_{2}$ and $(\mathbf{w}_{5})_{3}$ we compute $(\mathbf{w}_{5}) \times (\mathbf{w}_{5})$ and compare before and after splitting. After canceling the contributions we have just computed, we find:
\begin{equation} \label{6ieqn}
    (\mathbf{w}_{5})_{2} \times (\mathbf{w}_{5})_{2} = (\mathbf{w}_{5})_{2} \times (\mathbf{w}_{5})_{3} = (\mathbf{w}_{5})_{3} \times (\mathbf{w}_{5})_{2} = (\mathbf{w}_{5})_{3} \times (\mathbf{w}_{5})_{3} = 2(0) + (\mathbf{w}_{5})_{2} + (\mathbf{w}_{5})_{3}.
\end{equation}
Now, $(\mathbf{w}_{5})_{2}$ and $(\mathbf{w}_{5})_{3}$ must be mutual conjugates, since otherwise we obtain $(\mathbf{w}_{5})_{2} \times (\mathbf{w}_{5})_{3} = (\mathbf{w}_{5})_{3}$ or $(\mathbf{w}_{5})_{2} \times (\mathbf{w}_{5})_{3} = (\mathbf{w}_{5})_{2}$, both of which lead to the incorrect conclusion that $(\mathbf{w}_{5})_{2}$ or $(\mathbf{w}_{5})_{3}$ are the identity line. Overall, we find that the only consistent answer with \eqref{6ieqn} is
\begin{equation}
    (\mathbf{w}_{5})_{2} \times (\mathbf{w}_{5})_{2} = (\mathbf{w}_{5})_{3}, \quad (\mathbf{w}_{5})_{3} \times (\mathbf{w}_{5})_{3} = (\mathbf{w}_{5})_{2}, \quad (\mathbf{w}_{5})_{2} \times (\mathbf{w}_{5})_{3} = (\mathbf{w}_{5})_{3} \times (\mathbf{w}_{5})_{2} = (0),
\end{equation}
which is precisely the correct answer to reproduce the $\mathbb{Z}_{3}$ subfusion ring of $\mathrm{SU}(6)_{-1}$. Finally, considering $(3\mathbf{w}_{6}) \times (\mathbf{w}_{5}) = (\mathbf{w}_{1} + \mathbf{w}_{6})$, we find that $(\mathbf{w}_{1} + \mathbf{w}_{6})_{2} = (3\mathbf{w}_{6})_{1} \times (\mathbf{w}_{5})_{2}$ and $(\mathbf{w}_{1} + \mathbf{w}_{6})_{3} = (3\mathbf{w}_{6})_{1} \times (\mathbf{w}_{5})_{2}$ which reproduces the whole fusion ring of $\mathrm{SU}(6)_{-1}$.

\subsection*{Other Examples}

We finish this section by mentioning a few examples of Level-Rank dualities that we quote without verifying them in detail. For instance, a first example we may verify corresponds to entry number 36 in \cite{Schellekens:1992db}. Specifically:
\begin{equation}
    \frac{\mathrm{SU}(9)_{2}}{\mathbb{Z}_{3}} \cong \mathrm{F}_{4,-2} -48 \CSgrav,
\end{equation}
where the $\mathbb{Z}_{3}$ corresponds to the $\mathbb{Z}_{3}$ subgroup of the $\mathbb{Z}_{9}$ center.

Two more examples where it is straightforward to check that non-abelian condensation is required for the Level-Rank dualities to work are
\begin{equation}
    \mathrm{USp}(14)_{2}/\mathcal{A} \cong \mathrm{SU}(4)_{-1} - 48 \CSgrav,
\end{equation}
with the algebra
\begin{equation}
    \mathcal{A} = (0) + (\mathbf{w}_{1} + \mathbf{w}_{7}) + (2\mathbf{w}_{5}) + (\mathbf{w}_{2} + \mathbf{w}_{4}).
\end{equation}
and
\begin{equation}
\mathrm{F}_{4,6}/{\cal B}= \mathrm{SU}(3)_{-2} - 48 \CSgrav,
\end{equation}
with the algebra
\begin{equation}
    \mathcal{B} = (0) + (4 \mathbf{w}_{4}) + (3 \mathbf{w}_{3}) + (\mathbf{w}_{1} + \mathbf{w}_{2}).
\end{equation}
It is clear that both $\mathcal{A}$ and $\mathcal{B}$ pass the quantum dimension constraint \eqref{quantumdimensionconstraint}.

\section{$\mathrm{Spin}(n^{2})_{2}$ as a Dihedral Group Gauge Theory }\label{sec:dihedral}

Now that we have given concrete examples of dualities originating from holomorphic CFTs, we look into some general patterns of holomorphic CFTs that may be determined by examining their symmetry structure in further detail. 

To motivate the results of this section, let us study the Chern-Simons theory $\mathrm{Spin}(25)_{2}$ (entry number 57 in \cite{Schellekens:1992db}), which is one of the simplest examples of holomorphic CFT at $c = 24$ that does not correspond to a Niemeier theory. As explained in Sections \ref{secintro} and \ref{SectionDMNO}, this fact guarantees the existence of a topological boundary condition for $\mathrm{Spin}(25)_{2}$ with Lagrangian algebra dictated by the partition function of the associated holomorphic CFT. It is straightforward to perform the non-abelian anyon condensation to determine a $\mathrm{Rep}(\mathrm{D}_{5})$ fusion ring at the topological boundary. A similar result may be derived by studying chiral subalgebras of $\mathrm{E}_{8,1}$, the unique holomorphic CFT at $c=8$. Indeed, there exists a non-minimal embedding of $\mathrm{Spin}(9)_{2}$ into $\mathrm{E}_{8,1}$ that implies similar results as the $\mathrm{Spin}(25)_{2}$ case, but with $\mathrm{D}_{5}$ replaced by $\mathrm{D}_{3}$.

In this section, we demonstrate that the previous examples generalize to an infinite pattern. Namely, that for odd $n$, the Chern-Simons theory $\mathrm{Spin}(n^{2})_{2}$ is equivalent to the Drinfeld center of the $\mathrm{D}_{n}$ or $\mathrm{Rep}(\mathrm{D}_{n})$ fusion category symmetries; i.e. it is equivalent to $\mathrm{D}_{n}$ gauge theory for a specific value of the Dijkgraaf-Witten twist \cite{Dijkgraaf:1989pz}.

To see this, note the following:
\begin{equation}\label{chargeimport}
    \frac{\mathrm{Spin}(n^{2})_{2}}{\mathbb{Z}_{2}} \cong \mathrm{SO}(n^{2})_{2} \cong \mathrm{SU}(n^{2})_{1}.
\end{equation}
Conversely, we recover $\mathrm{Spin}(n^{2})_{2}$ from $\mathrm{SU}(n^{2})_{1}$ by gauging the zero-form charge-conjugation symmetry. Thus, we will show the desired statement by exhibiting $\mathrm{SU}(n^{2})_{1}$ in an appropriate abelian Chern-Simons form that we know gives rise to $\mathrm{D}_{n}$ gauge theory after gauging charge conjugation. 

To start, recall that any abelian MTC can be cast as an abelian Chern-Simons theory:
\begin{equation}
    \mathcal{L} = \frac{1}{4\pi} \sum^{n}_{i,j=1} A_{i} K_{ij} dA_{j}.
\end{equation}
The theory is gauge invariant when the matrix $K$ is symmetric and valued in integers $K \in \mathbb{Z}^{n \times n}$. The theory is bosonic if and only if the diagonal components are all even. If at least one diagonal component is odd, the theory is spin. The chiral central charge is given by the signature of the $K$-matrix: $c = \mathrm{sgn}(K)$.

The spectrum of line operators can be organized in terms of a vector $\vec{\alpha} \in \mathbb{Z}^{n}$ corresponding to the Wilson lines $W_{\vec{\alpha}}(\gamma) = \exp{i  \alpha^{i} \int_{\gamma} A_{i}}$, and their fusion is given by the standard addition of $\mathbb{Z}^{n}$ vectors. Lines that differ by an element lying in the image of $K$ are identified with each other $\vec{\alpha} \sim \vec{\alpha} + K \vec{\beta}$, and thus the set of independent lines is finite and labeled in the lattice $\mathbb{Z}^{n}/K\mathbb{Z}^{n}$.

The topological spins of the lines are given by:
\begin{equation}
    \theta_{\vec{\alpha}} = \exp(2 \pi i h_{\vec{\alpha}}), \quad  h_{\vec{\alpha}} = \frac{1}{2} \vec{\alpha}^{t} K^{-1} \vec{\alpha} \, .
\end{equation}
The braiding matrix is fully determined by the topological spins of the anyons and their fusion:
\begin{equation}
    B(\vec{\alpha}, \vec{\beta}) = \frac{\theta_{\vec{\alpha} + \vec{\beta}}}{\theta_{\vec{\alpha}} \, \theta_{\vec{\beta}}} \, .
\end{equation}

We claim that we can characterize $\mathrm{SU}(n^{2})_{1}$ in terms of an abelian Chern-Simons theory with the following $K$-matrix (up to a gravitational Chern-Simons term):
\begin{equation} \label{kmatrix1}
   K =  
      \begin{pmatrix}
            4 & n \\
            n & 0 
        \end{pmatrix}~.
\end{equation}
To see that $\mathrm{SU}(n^{2})_{1}$ can be written in this form, recall that an abelian MTC is fully characterized by the topological spin of the anyons and their fusion rules. Since $n$ is odd, it is straightforward to see that we can label the independent anyons as $\vec{\alpha}^{T} = (0,\alpha)$, with $\alpha$ an integer mod $n^{2}$. The fusion ring is clearly $\mathbb{Z}_{n^{2}}$, and the topological spins of the lines are:
\begin{equation}
    \theta_{\alpha} = \exp( - 2 \pi i \frac{2 \alpha^{2}}{n^{2}}).
\end{equation}
Meanwhile, the conformal weights of $\mathrm{SU}(n^{2})_{1}$ are given by (see Appendix \ref{SummaryAnyonDataAppendix}):
\begin{equation}
    h_{i} = \frac{i(n^{2}-i)}{2n^{2}}, \quad i=0, 1, \ldots, n^{2}-1.
\end{equation}
Then, it is easy to verify that the line $\alpha = (n^{2} + 1)/2$ maps to the $i=1$ line in the standard description of $\mathrm{SU}(n^{2})_{1}$. Since the fusion rules and the spins of the generators match,  we see that the $K$-matrix \eqref{kmatrix1} indeed describes $\mathrm{SU}(n^{2})_{1}$. 

\begin{figure}[t]
\centering
        \includegraphics[scale=1.4]{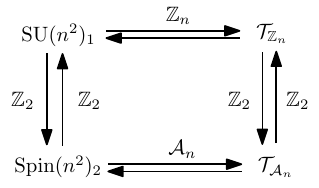} 
        \caption{Gauging relations between $\mathrm{SU}(n^{2})_{1}$ and $\mathrm{Spin}(n^{2})_{2}$ for $n$ odd, and the holomorphic CFTs $\mathcal{T}_{\mathbb{Z}_{n}}$ and $\mathcal{T}_{\mathcal{A}_{n}}$ defined by gauging the $\mathbb{Z}_{n}$ subgroup of the $\mathbb{Z}_{n^{2}}$ center symmetry of $\mathrm{SU}(n^{2})_{1}$, and by gauging an algebra $\mathcal{A}_{n}$ in $\mathrm{Spin}(n^{2})_{2}$, respectively. Studying the associated topological boundaries (see Figure \ref{HoloCFTPartitionFunction}), one can verify that $\mathcal{T}_{\mathbb{Z}_{n}}$ comes equipped with a $\mathbb{Z}_{n}$ symmetry, while $\mathcal{T}_{\mathcal{A}_{n}}$ comes equipped with a $\mathrm{Rep}(\mathrm{D}_{n})$ symmetry. Special cases of the diagram arise for $n = 3$ where $\mathcal{T}_{\mathbb{Z}_{3}} = \mathcal{T}_{\mathcal{A}_{3}} = \mathrm{E}_{8,1}$, and for $n=5$ where $\mathcal{T}_{\mathbb{Z}_{5}} = \mathrm{Sch}_{67}$ and $\mathcal{T}_{\mathcal{A}_{5}} = \mathrm{Sch}_{57}$.}  \label{CommutativeDiagramgeneraln}
\end{figure}

To connect this derivation to finite gauge theory, notice that $\mathrm{U(1)\times U(1)}$ Chern-Simons theory with level matrix \eqref{kmatrix1}  is equivalent to a twisted $\mathbb{Z}_{n}$ gauge theory.  Indeed, the off-diagonal terms truncate the gauge group to $\mathbb{Z}_{n}$ while the diagonal term specifies the specific twist. Finally, we gauge charge conjugation.  In the $\mathbb{Z}_{n}$ gauge theory, this acts as an outer automorphism of the gauge group and promotes the gauge group to the Dihedral group $\mathrm{D}_{n}$ \cite{Cordova:2024jlk, Cordova:2024mqg}.  In the Chern-Simons description this converts the gauge group to $\mathrm{Spin}(n^{2})_{2}$ as explained below \eqref{chargeimport}.  Thus we find:
\begin{equation}
    \mathrm{Spin}(n^{2})_{2} \cong \mathrm{Twisted}  \ \mathrm{D}_{n} \ \mathrm{Gauge \ Theory}~,
\end{equation}
generalizing the sporadic examples at $c=8$ and $c=24$ discussed above to any odd $n.$  A summary of the general situation is depicted in Figure \ref{CommutativeDiagramgeneraln}, showing the relation between $\mathrm{SU}(n^{2})_{1}$, $\mathrm{Spin}(n^{2})_{2}$, and the holomorphic CFTs they define after gauging.

\section{The Center of $\mathbb{Z}_{2n+1}$ TY as $\mathrm{Spin}(2n + 1)_{2} \times \mathrm{SU}(2n+1)_{-1}$ CS Theory } \label{section5}

In this section we derive another infinite pattern of dualities by extrapolating from specific holomorphic CFTs.

Our analysis begins by noticing that there exist two different holomorphic CFTs at $c = 24$ that can in principle be characterized in terms of a $\mathrm{Spin}(13)_{2} \times \mathrm{Spin}(13)_{2}$ chiral algebra (see Table \ref{Spinoddlv2table} to for the spectrum of $\mathrm{Spin}(13)_{2}$). One of them corresponds to entry number 41 in \cite{Schellekens:1992db}. The partition function is:
\begin{align} \label{Sch41Z}
    & Z_{\mathrm{Sch}_{41}}(\tau) = (\chi^{\mathrm{B}_{6,2}}_{0} \chi^{\mathrm{B}_{6,2}}_{0} + \chi^{\mathrm{B}_{6,2}}_{\eta} \chi^{\mathrm{B}_{6,2}}_{\eta} + \chi^{\mathrm{B}_{6,2}}_{A} \chi^{\mathrm{B}_{6,2}}_{B} + \chi^{\mathrm{B}_{6,2}}_{B} \chi^{\mathrm{B}_{6,2}}_{A} + \nonumber \\[0.25cm] & \chi^{\mathrm{B}_{6,2}}_{\ell_{1}} \chi^{\mathrm{B}_{6,2}}_{\ell_{5}} + \chi^{\mathrm{B}_{6,2}}_{\ell_{5}} \chi^{\mathrm{B}_{6,2}}_{\ell_{1}} + \chi^{\mathrm{B}_{6,2}}_{\ell_{2}} \chi^{\mathrm{B}_{6,2}}_{\ell_{3}} + \chi^{\mathrm{B}_{6,2}}_{\ell_{3}} \chi^{\mathrm{B}_{6,2}}_{\ell_{2}} + \chi^{\mathrm{B}_{6,2}}_{\ell_{4}} \chi^{\mathrm{B}_{6,2}}_{\ell_{6}} + \chi^{\mathrm{B}_{6,2}}_{\ell_{6}} \chi^{\mathrm{B}_{6,2}}_{\ell_{4}})(\tau).
\end{align}
As discussed in Section \ref{SectionDMNO}, this expression corresponds to the combination of anyons generating the Lagrangian algebra in $\mathrm{Spin}(13)_{2} \times \mathrm{Spin}(13)_{2}$ required to construct the holomorphic CFT from the 3D TQFT viewpoint.

Meanwhile, entry number 60 in \cite{Schellekens:1992db} is based on the chiral algebra $\mathrm{SU(13)_{1} \times SU(13)_{1}}$. The partition function is:
\begin{align}
    & Z_{\mathrm{Sch}_{60}}(\tau) = \chi^{\mathrm{A}_{12,1}}_{0}(\tau) \chi^{\mathrm{A}_{12,1}}_{0}(\tau) + \sum_{i=1}^{12} \chi^{\mathrm{A}_{12,1}}_{i}(\tau) \chi^{\mathrm{A}_{12,1}}_{5i}(\tau), 
\end{align}
where the labeling of the primaries follows the nomenclature of Appendix \ref{SummaryAnyonDataAppendix}. Notice in particular that the labels of the characters above are defined mod 13 (e.g. $\chi^{\mathrm{A}_{12,1}}_{15}(\tau) \coloneqq \chi^{\mathrm{A}_{12,1}}_{2}(\tau))$. It is also clear that there exists an obvious $\mathbb{Z}_{13}$ symmetry assigning a $\mathbb{Z}_{13}$ charge to the primaries of the theory. 

There is, however, a conformal embedding of affine Lie algebras $\mathrm{B}_{6,2} \hookrightarrow \mathrm{A}_{12,1}$ \cite{PhysRevD.34.3092, ALEXANDERBAIS1987561}, from which the $\mathrm{SU}(13)_{1}$ characters can be expressed in terms of those of $\mathrm{Spin}(13)_{2}$ using the branching rules of the embedding. The partition function written in terms of $\mathrm{Spin}(13)_{2}$ characters is:
\begin{align} \label{Sch60Z}
    & Z_{\mathrm{Sch}_{60}}(\tau) = ((\chi^{\mathrm{B}_{6,2}}_{0} + \chi^{\mathrm{B}_{6,2}}_{\eta})(\chi^{\mathrm{B}_{6,2}}_{0} + \chi^{\mathrm{B}_{6,2}}_{\eta}) + \nonumber \\[0.25cm] & 2(\chi^{\mathrm{B}_{6,2}}_{\ell_{1}} \chi^{\mathrm{B}_{6,2}}_{\ell_{5}} + \chi^{\mathrm{B}_{6,2}}_{\ell_{5}} \chi^{\mathrm{B}_{6,2}}_{\ell_{1}} + \chi^{\mathrm{B}_{6,2}}_{\ell_{2}} \chi^{\mathrm{B}_{6,2}}_{\ell_{3}} + \chi^{\mathrm{B}_{6,2}}_{\ell_{3}} \chi^{\mathrm{B}_{6,2}}_{\ell_{2}} + \chi^{\mathrm{B}_{6,2}}_{\ell_{4}} \chi^{\mathrm{B}_{6,2}}_{\ell_{6}} + \chi^{\mathrm{B}_{6,2}}_{\ell_{6}} \chi^{\mathrm{B}_{6,2}}_{\ell_{4}}))(\tau).
\end{align}
Clearly, in terms of the 3D bulk construction of holomorphic CFTs, the difference between the holomorphic CFTs with partition functions \eqref{Sch41Z} and \eqref{Sch60Z} arises from different choices of topological boundary condition (see Figure \ref{HoloCFTPartitionFunction}). As mentioned in Section \ref{secintro}, this means the fusion category of symmetries made manifest in each topological boundary is different in each case. (This is also straightforward to see in the previous examples by direct calculation.)

As reviewed in Section \ref{secintro}, all holomorphic CFTs at $c=24$ can be obtained from each other by (non-invertible) gauging \cite{van_Ekeren_2017, Hohn:2017dsm, Hohn:2023auw, vanEkeren2021, Moller2023, hohn2022systematic, moller2024geometric}. In particular, the two holomorphic CFTs above are related by a $\mathbb{Z}_{2}$ gauging. The relationship between the two distinct sets of symmetries is interesting. For instance, performing the anyon condensation, one can convince oneself that entry 41 in \cite{Schellekens:1992db} has a $\mathrm{Spin}(13)_{2}$ fusion category symmetry. Indeed, the Lagrangian algebra that follows from equation \eqref{Sch41Z} implies that $\mathrm{Spin}(13)_{2}$ is time-reversal symmetric (recall the discussion in Section \ref{SectionDMNO}). Thus, the fusion category at the topological boundary is essentially the same as that of the diagonal Lagrangian algebra in $\mathrm{Spin}(13)_{2} \times \mathrm{Spin}(13)_{-2}$, i.e.\ $\mathrm{Spin}(13)_{2}$. In turn, $\mathrm{Spin}(13)_{2}$ contains a $\mathrm{Rep}(\mathrm{D}_{13})$ subcategory \cite{Ardonne_2016}, which is expected from the $\mathbb{Z}_{13}$ symmetry of entry 60 of \cite{Schellekens:1992db} and the $\mathbb{Z}_{2}$ gauging. More intriguing is the fact that $\mathrm{Spin}(13)_{2}$ contains two additional lines outside $\mathrm{Rep}(\mathrm{D}_{13})$ that are not accounted for by the $\mathbb{Z}_{13}$ symmetry of entry 41 of \cite{Schellekens:1992db}. The fusion rules are similar to those of a duality defect, but where the algebra one gauges is that of a non-invertible symmetry. See Table \ref{Spinoddlv2table} for the spectrum and equation \eqref{eqn1} for the relevant fusion rules.

In the following, we will use the preceding concrete case to guide us through a generalization to an infinite family of examples. First, we observe that the Chern-Simons theory $\mathrm{Spin}(2n + 1)_{2} \times \mathrm{SU}(2n+1)_{-1}$ for positive integer $n$ is equivalent to the Drinfeld center of the $\mathbb{Z}_{2n+1}$ Tambara-Yamagami fusion category for a specific value of the bicharacter and Frobenius-Schur indicator. This can be shown by direct analysis of the anyon condensation, or by arguments based on manipulations of the action.  This equivalence has also been noted in the mathematics literature \cite{Ardonne_2016, Ardonne_2021, Evans:2023vns}. Once we have this result, it will be straightforward to perform a few topological manipulations to reproduce both the observations above and determine a generalization to an appropriate infinite family. 

\begin{table}[t]
\centering
\begin{tabular}[h]{|p{2cm}|p{2.8cm}|p{3.0cm}|p{3.0cm}| }
\hline 
\multicolumn{4}{|c|}{$\mathrm{Spin}(2n+1)_{2}$} \\
\hline
Line label & Highest Weight & Quantum Dim. & Conf. Weight \\
\hline
0 & (0) & $d_{0} = 1$ & $h_{0} = 0$ \\
$\eta$ & $(2 \mathbf{w}_{1})$ & $d_{\eta} = 1$ & $h_{\eta} = 1$ \\
A & $(\mathbf{w}_{\sigma})$ & $d_{A} = \sqrt{2n + 1}$ & $h_{A} = n/8$ \\
B & $(\mathbf{w}_{1} + \mathbf{w}_{\sigma})$ & $d_{B} = \sqrt{2n + 1}$ & $h_{B} = n/8+1/2$ \\
$\ell_{i}$ & $(\mathbf{w}_{i})$  & $d_{\ell_{i}} = 2$ & $h_{\ell_{i}} = \frac{i(2n+1-i)}{2(2n+1)}$ \\
$\ell_{n}$ & $(2\mathbf{w}_{\sigma})$ & $d_{\ell_{n}} = 2$ & $h_{\ell_{n}} = \frac{n(n+1)}{2(2n+1)}$ \\
\hline
\end{tabular}
\caption{Data of $\mathrm{Spin}(2n+1)_{2}$. The label $i$ runs through $i=1, \ldots, n-1$.}  \label{Spinoddlv2table}
\end{table}

\subsection*{Condensation Analysis}

We begin from the following observation:
\begin{equation} \label{SpinToSUEmbedding}
    \frac{\mathrm{Spin}(2n + 1)_{2}}{\mathbb{Z}_{2}} = \mathrm{SO}(2n+1)_{2} = \mathrm{SU}(2n+1)_{1},
\end{equation}
which follows from the existence of an embedding of chiral algebras $\mathrm{B}_{n,2} \hookrightarrow \mathrm{A}_{2n,1}$ (see e.g. the list of conformal embeddings in \cite{davydov2013witt}).

Then, we know that there exists a topological interface between $\mathrm{Spin}(2n + 1)_{2}$ and $\mathrm{SU}(2n+1)_{1}$, and thus $\mathrm{Spin}(2n + 1)_{2} \times \mathrm{SU}(2n+1)_{-1}$ admits a topological boundary condition. To deduce which fusion category lives at this boundary, we can analyze the interface obtained by gauging the $\mathbb{Z}_{2}$ symmetry in $\mathrm{Spin}(2n + 1)_{2}$ using the tools summarized in Appendix \ref{GaugingReviewAppendix}.

The spectrum of $\mathrm{Spin}(2n + 1)_{2}$ consists of $n+4$ anyons, whose data is summarized in Table \ref{Spinoddlv2table}. The relevant fusion rules of $\mathrm{Spin}(2n + 1)_{2}$ are given by:
\begin{equation} \label{eqn1}
    A \times A = B \times B = 1 + \sum_{s=1}^{n} \ell_{s}, \quad A \times B = \eta + \sum_{s=1}^{n} \ell_{s}, \quad \eta \times A = B
\end{equation}
\begin{equation} \label{eqn2}
    \eta \times \ell_{s} = \ell_{s}, \quad \ell_{s} \times \ell_{s} = 1 + \eta + \ldots, \quad s = 1, \ldots, n.
\end{equation}

Since we are just gauging $\eta$, the algebra is simply $\mathcal{A} = 1 + \eta$. Then, the lines $\ell_{i}$ do not confine and split into two lines of quantum dimension one. These are the conjugate pairs in $\mathrm{SU}(2n+1)_{1}$. Meanwhile, condensing $\eta \rightarrow 1$, implies that $A$ and $B$ are identified. From the fusion rules we also see that $A$ and $B$ do not split, and from the topological spins we see that they confine at the interface between $\mathrm{Spin}(2n + 1)_{2}$ and $\mathrm{SU}(2n+1)_{1}$ and do not descend into genuine line operators of $\mathrm{SU}(2n+1)_{1}$. Since the lines $\ell_{i}$ descend into the abelian lines of $\mathrm{SU}(2n+1)_{1}$, we see from \eqref{eqn1} that the corresponding fusion ring is that of $\mathbb{Z}_{2n+1}$ Tambara-Yamagami.

\subsection*{Action Principle Analysis}

In this version of the analysis we start from $\mathrm{SU}(2n+1)_{1} \times \mathrm{SU}(2n+1)_{-1}$ Chern-Simons theory. Gauging the charge-conjugation zero-form symmetry on the left factor takes us to $\mathrm{Spin}(2n+1)_{2} \times \mathrm{SU}(2n+1)_{-1}$. We claim that $\mathrm{SU}(2n+1)_{1} \times \mathrm{SU}(2n+1)_{-1}$ is equivalent to untwisted $\mathbb{Z}_{2n+1}$ gauge theory. Gauging charge conjugation in the standard formulation of $\mathrm{SU}(2n+1)_{1} \times \mathrm{SU}(2n+1)_{-1}$ corresponds then to gauging a zero-form symmetry that exchanges the electric and magnetic charges in the standard formulation of untwisted $\mathbb{Z}_{2n+1}$ gauge theory, thus making manifest that the topological boundary will now host a $\mathbb{Z}_{2n+1}$ Tambara-Yamagami fusion category. This establishes $\mathrm{Spin}(2n+1)_{2} \times \mathrm{SU}(2n+1)_{-1}$ Chern-Simons theory as its Drinfeld center.

In the standard formulation of $\mathrm{SU}(2n+1)_{1} \times \mathrm{SU}(2n+1)_{-1}$ Chern-Simons theory, the spins of the lines are given by (see Appendix \ref{SummaryAnyonDataAppendix}):
\begin{equation}
    h_{(i,j)} = \frac{i(2n+1-i)}{2(2n+1)} - \frac{j(2n+1-j)}{2(2n+1)},
\end{equation}
and the fusion rules are $\mathbb{Z}_{2n+1} \times \mathbb{Z}_{2n+1}$. Meanwhile, untwisted $\mathbb{Z}_{2n+1}$ gauge theory is described by a $K$-matrix
\begin{equation} \label{kmatrix2}
   K =  
      \begin{pmatrix}
            0 & 2n+1 \\
            2n+1 & 0 
        \end{pmatrix},
\end{equation}
with topological spins given by
\begin{equation}
    \theta_{[\alpha,\beta]} = \frac{\alpha \beta}{2n+1}.
\end{equation}
The fusion rules are easily derived from $K$-matrix identifications and are also $\mathbb{Z}_{2n+1} \times \mathbb{Z}_{2n+1}$. Throughout this section and only in this section we use $(i,j)$ to denote the lines in $\mathrm{SU}(2n+1)_{1} \times \mathrm{SU}(2n+1)_{-1}$, while we use square brackets $[\alpha, \beta]$ to denote lines in the untwisted $\mathbb{Z}_{2n+1}$ gauge theory formulation of the theory.

To match the two descriptions we use again that an abelian MTC is fully determined by the topological spin of the lines and their fusion rules. It is straightforward to match:
\begin{equation}
    \Big( (n+1)  \alpha - \beta , \quad (n+1) \alpha + \beta \Big) \longleftrightarrow [\alpha, \beta],
\end{equation}
or inversely:
\begin{equation}
    ( i, j) \longleftrightarrow \Big[ i + j, \quad  (n+1) (j-i) \Big],
\end{equation}
which has the same topological spins and fusion rules on both sides.

Now that we have shown that these two descriptions match, we must check that the charge-conjugation zero form symmetry $\mathcal{C}_{L}$ of the left factor in the $\mathrm{SU}(2n+1)_{1} \times \mathrm{SU}(2n+1)_{-1}$ description maps to an electric-magnetic duality symmetry in the untwisted $\mathbb{Z}_{2n+1}$ gauge theory description. This is, we must find $\alpha'$ and $\beta'$ such that 
\begin{align}
     & \mathcal{C}_{L}\Bigg(  \Big( (n+1) \alpha - \beta , \quad (n+1) \alpha + \beta \Big) \Bigg) = \nonumber \\[0.4cm]
    & \Big( - (n+1) \alpha + \beta , \quad (n+1) \alpha + \beta \Big) = \Big( (n+1) \alpha' - \beta' , \quad (n+1) \alpha' + \beta' \Big).
\end{align}
It is easy to see then that $\alpha' = 2 \beta$ and $\beta' = (n+1) \alpha$, and thus we deduce that on the $\mathbb{Z}_{2n+1}$ gauge theory formulation of the theory the zero-form symmetry $\mathcal{C}_{L}$ acts as
\begin{equation}
    \mathcal{C}_{L} \big[ [\alpha, \beta] \big] = [2 \beta, (n+1) \alpha],
\end{equation}
indeed exchanging electric and magnetic charges.

\subsection*{Relationship with Time-Reversal Symmetry and Holomorphic CFT}

To relate the statement that $\mathrm{Spin}(2n + 1)_{2} \times \mathrm{SU}(2n+1)_{-1}$ is the center of $\mathbb{Z}_{2n+1}$ TY to Holomorphic CFT, let us notice that $\mathrm{SU}(2n+1)_{1}$ is a time-reversal invariant theory (as a bosonic theory) whenever $k \coloneqq 2n+1$ is such that $-1$ is a quadratic residue modulo $k$.\footnote{This is the same condition derived in \cite{Delmastro:2019vnj} for $\mathrm{U}(1)_{k} \equiv \mathrm{SU}(k)_{-1}$ to be time-reversal invariant as a spin theory. } 

Assuming $-1$ is a quadratic residue modulo $k$, we can reverse the level of the second factor in $\mathrm{Spin}(2n + 1)_{2} \times \mathrm{SU}(2n+1)_{-1}$ at the cost of a gravitational Chern-Simons term. Thus, we see that $\mathrm{Spin}(k)_{2} \times \mathrm{SU}(k)_{1}$ defines an holomorphic CFT whenever $-1$ is a quadratic residue modulo $k$. In other words, there exists an algebra $\mathcal{A}_{k}$ such that
\begin{equation}
     \frac{\mathrm{Spin}(k)_{2} \times \mathrm{SU}(k)_{1}}{\mathcal{A}_{k}} \cong  -4(k-1)\CSgrav,
\end{equation}
when $-1$ is a quadratic residue modulo $k$. Analyzing the splittings in \eqref{eqn1} and \eqref{eqn2}, and keeping track of the time-reversal transformation, it is straightforward to see that $\mathcal{A}_{k}$ is given by:
\begin{align}
    \mathcal{A}_{k} = \big(0, 0 \big) + \big((2\mathbf{w}_{1}), 0 \big) &+ \sum_{i=1}^{\frac{(k-3)}{2}}\big[ \big((\mathbf{w}_{i}), q(i)\big) + \big((\mathbf{w}_{i}), q(k-i)\big) \big] \nonumber \\[0.25cm] &+ \big((2\mathbf{w}_{\sigma}), q \bigg( \frac{k-1}{2} \bigg) ) + \big((2\mathbf{w}_{\sigma}), q \bigg(\frac{k+1}{2} \bigg) \big),
\end{align}
where $q(\sigma)$ stands for the time-reversal permutation that brings the line $\sigma$ mod $k$ in $\mathrm{SU}(k)_{1}$ to the line $q \sigma$ mod $k$ (see appendices \ref{SummaryAnyonDataAppendix} and \ref{SU(k)1TimeReversalAppendix}). The fusion category living at the topological boundary separating $\mathrm{Spin}(k)_{2} \times \mathrm{SU}(k)_{1}$ from the gravitational Chern-Simons term is the $\mathbb{Z}_{k}$ Tambara-Yamagami fusion category we found in the prior subsections. For $k=13$, it is straightforward to verify that the resulting holomorphic CFT corresponds to entry 60 of \cite{Schellekens:1992db} by comparing the algebra with the partition function given by \eqref{Sch60Z}. However, with only one of the $\mathrm{SU}(13)_{1}$ factors decomposed in terms of $\mathrm{Spin}(13)_{2}$ characters. The construction thus far clearly generalizes the discussion above from $k=13$ to any $k$ for which $-1$ is a quadratic residue modulo $k$.

We can now gauge charge conjugation in the $\mathrm{Spin}(k)_{2} \times \mathrm{SU}(k)_{1}$ TQFT. Since charge conjugation acts trivially on $\mathrm{Spin}(k)_{2}$ (all lines are self-conjugate), this is equivalent to gauging charge conjugation on the $\mathrm{SU}(k)_{1}$ factor alone. This gives a second factor of $\mathrm{Spin}(k)_{2}$,\footnote{It is straightforward to verify that gauging charge conjugation is the inverse of the $\mathbb{Z}_{2}$ gauging in \eqref{SpinToSUEmbedding}. See \cite{Barkeshli:2014cna} for a explicit calculation showing this fact.} and we obtain a TQFT $\mathrm{Spin}(k)_{2} \times \mathrm{Spin}(k)_{2}$ separated by a topological boundary to the gravitational Chern-Simons term $-4(k-1)\CSgrav$.

Since in bulk we are gauging charge-conjugation, the latter topological interface will now host a fusion category symmetry corresponding to a gauging of charge conjugation of the $\mathbb{Z}_{k}$ Tambara-Yamagami fusion category previously found.

\begin{figure}[t]
\centering
        \includegraphics[scale=1.5]{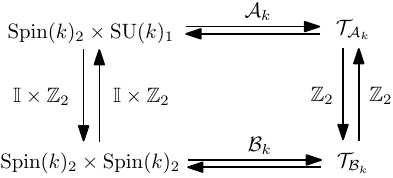} 
        \caption{When $-1$ is a quadratic residue modulo $k$, we can gauge an algebra $\mathcal{A}_{k}$ in $\mathrm{Spin}(k)_{2} \times \mathrm{SU}(k)_{1}$ to find a holomorphic CFT $\mathcal{T}_{\mathcal{A}_{k}}$ equipped with a $\mathbb{Z}_{k}$ Tambara-Yamagami symmetry. After gauging charge conjugation, this symmetry becomes the $\mathbb{Z}_{2}$-equivariantization of $\mathbb{Z}_{k}$ Tambara-Yamagami. This is equivalent to the $\mathrm{Spin}(k)_{2}$ fusion category, which now appears as the symmetry of a holomorphic CFT $\mathcal{T}_{\mathcal{B}_{k}}$ that arises after gauging an algebra $\mathcal{B}_{k}$ in $\mathrm{Spin}(k)_{2} \times \mathrm{Spin}(k)_{2}$. Special cases of the diagram arise for $k=5$ where $\mathcal{T}_{\mathcal{A}_{5}} = \mathcal{T}_{\mathcal{B}_{5}} = \mathrm{E}_{8,1}$, and for $k=13$ where $\mathcal{T}_{\mathcal{A}_{13}} = \mathrm{Sch}_{60}$ and $\mathcal{T}_{\mathcal{B}_{13}} = \mathrm{Sch}_{41}$.} \label{CommutativeDiagramgeneralk}
\end{figure}

This construction corresponds to the $\mathbb{Z}_{2}$-equivariantization of $\mathbb{Z}_{k}$ Tambara-Yamagami, which has been studied e.g. in \cite{Ardonne_2016, Ardonne_2021, Evans:2023vns}. We follow \cite{Evans:2023vns}. Briefly, the simple objects in the $\IZ_2$-equivariantization of a Tambara-Yamagami category based on a discrete group $G$ has simple objects $\beta_h^t$  with quantum  dimension $1$, $\sigma_g$ for any $g \in G$ with $g^{2} \ne 1$ with quantum dimension $2$ and $\rho^t$ with quantum dimension $\sqrt{|G|}$. Here $t = \pm 1$, $h$ are elements of $G$ with $h^{2}=1$, and $\sigma_g=\sigma_{-g}$. The fusion rules are given in equation (5.26) of \cite{Evans:2023vns}.\footnote{Notice that the second equation in the first row of (5.26) of  \cite{Evans:2023vns} should read $[\rho^{t}] [\sigma_{g}] = [\rho^{t}] + [\rho^{-t}]$. We would like to thank D. E. Evans for confirming this to us.}

Specializing to $G=\IZ_{k}$ with $k$ odd and writing the group additively with elements $\{0,1,2, \cdots k-1 \}$, the only element $h$ with $2h=0$ is $h=0$. Thus, we have simples $\beta^{\pm}$ with quantum dimension $1$, $\sigma_1, \sigma_2, \cdots \sigma_{(k-1)/2}$ with
quantum dimension $2$ and $\rho^\pm$ with quantum dimensions $\sqrt{k}$. Comparing with Table \ref{Spinoddlv2table}, and following \cite{Evans:2023vns}, we observe a exact match with the spectrum and fusion rules of $\mathrm{Spin}(k)_{2}$. Thus, we conclude that $\mathrm{Spin}(k)_{2} \times \mathrm{Spin}(k)_{2}$ admits a gauging by an algebra $\mathcal{B}_{k}$ that results in a holomorphic CFT equipped with a $\mathrm{Spin}(k)_{2}$ fusion category. The additional lines $\rho^\pm$ in $\mathrm{Spin}(k)_{2}$ outside the $\mathrm{Rep}(\mathrm{D}_{k})$ subfusion ring arise due to the $\mathbb{Z}_{k}$ duality line that was present before gauging charge-conjugation, thus recovering and generalizing the observation made previously for $k=13$. A summary of the general situation is presented in Figure \ref{CommutativeDiagramgeneralk}.

\acknowledgments
We thank B. Rayhaun for helpful conversations. DGS is supported by the Society of Fellows at Harvard University. CC and DGS are supported by the Simons Collaboration on Global Categorical Symmetries, the US Department of Energy Grant 5-29073, and the Sloan Foundation. JH is supported by the National Science Foundation Grant PHY-2310635.

\appendix
\section{Schellekens' List of $c=24$ Holomorphic CFTs \cite{Schellekens:1992db}} \label{AppSchellekens}

\begin{table}[htbp]
\centering
\small
\setlength{\tabcolsep}{4pt}
\renewcommand{\arraystretch}{.98}

\begin{tabularx}{\textwidth}{|l|l|>{\raggedright\arraybackslash}X||l|l|>{\raggedright\arraybackslash}X|}
\hline
\multicolumn{1}{|c|}{No.} &
\multicolumn{1}{c|}{$\mathrm{dim}(V_{1})$} &
\multicolumn{1}{c||}{Spin-1 algebra} &
\multicolumn{1}{c|}{No.} &
\multicolumn{1}{c|}{$\mathrm{dim}(V_{1})$} &
\multicolumn{1}{c|}{Spin-1 algebra} \\
\hline\hline
0  & 0   & ---                                & 36 & 132  & $\mathrm{\mathrm{A}_{8,2} \mathrm{F}_{4,2}}$ \\
\hline
1  & 24  & $\mathrm{U}(1)^{24}$                        & 37 & 144  & $(\mathrm{A}_{4,1})^6$ \\
\hline
2  & 36  & $(\mathrm{A}_{1,4})^{12}$                   & 38 & 144  & $(\mathrm{C}_{4,1})^4$ \\
\hline
3  & 36  & $\mathrm{D}_{4,12}\mathrm{A}_{2,6}$                  & 39 & 144  & $\mathrm{D}_{6,2}\mathrm{C}_{4,1}(\mathrm{B}_{3,1})^2$ \\
\hline
4  & 36  & $\mathrm{C}_{4,10}$                         & 40 & 144  & $\mathrm{A}_{9,2}\mathrm{A}_{4,1}\mathrm{B}_{3,1}$ \\
\hline
5  & 48  & $(\mathrm{A}_{1,2})^{16}$                   & 41 & 156  & $(\mathrm{B}_{6,2})^2$ \\
\hline
6  & 48  & $(\mathrm{A}_{2,3})^6$                      & 42 & 168  & $(\mathrm{D}_{4,1})^6$ \\
\hline
7  & 48  & $(\mathrm{A}_{3,4})^3\mathrm{A}_{1,2}$               & 43 & 168  & $(\mathrm{A}_{5,1})^4\mathrm{D}_{4,1}$ \\
\hline
8  & 48  & $\mathrm{A}_{5,6}\mathrm{C}_{2,3}\mathrm{A}_{1,2}$            & 44 & 168  & $\mathrm{E}_{6,2}\mathrm{C}_{5,1}\mathrm{A}_{5,1}$ \\
\hline
9  & 48  & $(\mathrm{A}_{4,5})^2$                      & 45 & 168  & $\mathrm{E}_{7,3}\mathrm{A}_{5,1}$ \\
\hline
10 & 48  & $\mathrm{D}_{5,8}\mathrm{A}_{1,2}$                   & 46 & 192  & $(\mathrm{A}_{6,1})^4$ \\
\hline
11 & 48  & $\mathrm{A}_{6,7}$                          & 47 & 192  & $\mathrm{D}_{8,2}(\mathrm{B}_{4,1})^2$ \\
\hline
12 & 60  & $(\mathrm{C}_{2,2})^{6}$                    & 48 & 192  & $(\mathrm{C}_{6,1})^2\mathrm{B}_{4,1}$ \\
\hline
13 & 60  & $\mathrm{D}_{4,4}(\mathrm{A}_{2,2})^4$               & 49 & 216  & $(\mathrm{A}_{7,1})^2(\mathrm{D}_{5,1})^2$ \\
\hline
14 & 60  & $\mathrm{F}_{4,6}\mathrm{A}_{2,2}$                   & 50 & 216  & $\mathrm{D}_{9,2}\mathrm{A}_{7,1}$ \\
\hline
15 & 72  & $(\mathrm{A}_{1,1})^{24}$                   & 51 & 240  & $(\mathrm{A}_{8,1})^3$ \\
\hline
16 & 72  & $(\mathrm{A}_{3,2})^4(\mathrm{A}_{1,1})^4$           & 52 & 240  & $\mathrm{C}_{8,1}(\mathrm{F}_{4,1})^2$ \\
\hline
17 & 72  & $\mathrm{A}_{5,3}\mathrm{D}_{4,3}(\mathrm{A}_{1,1})^3$        & 53 & 240  & $\mathrm{E}_{7,2}\mathrm{B}_{5,1}\mathrm{F}_{4,1}$ \\
\hline
18 & 72  & $\mathrm{A}_{7,4}(\mathrm{A}_{1,1})^3$               & 54 & 264  & $(\mathrm{D}_{6,1})^4$ \\
\hline
19 & 72  & $\mathrm{D}_{5,4}\mathrm{C}_{3,2}(\mathrm{A}_{1,1})^2$        & 55 & 264  & $(\mathrm{A}_{9,1})^2\mathrm{D}_{6,1}$ \\
\hline
20 & 72  & $\mathrm{D}_{6,5}(\mathrm{A}_{1,1})^2$               & 56 & 288  & $\mathrm{C}_{10,1}\mathrm{B}_{6,1}$ \\
\hline
21 & 72  & $\mathrm{C}_{5,3} \mathrm{G}_{2,2} \mathrm{A}_{1,1}$            & 57 & 300  & $\mathrm{B}_{12,2}$ \\
\hline
22 & 84  & $\mathrm{C}_{4,2}(\mathrm{A}_{4,2})^2$               & 58 & 312  & $(\mathrm{E}_{6,1})^4$ \\
\hline
23 & 84  & $(\mathrm{B}_{3,2})^4$                      & 59 & 312  & $\mathrm{A}_{11,1}\mathrm{D}_{7,1}\mathrm{E}_{6,1}$ \\
\hline
24 & 96  & $(\mathrm{A}_{2,1})^{12}$                   & 60 & 336  & $(\mathrm{A}_{12,1})^2$ \\
\hline
25 & 96  & $(\mathrm{D}_{4,2})^2(\mathrm{C}_{2,1})^4$           & 61 & 360  & $(\mathrm{D}_{8,1})^3$ \\
\hline
26 & 96  & $(\mathrm{A}_{5,2})^2\mathrm{C}_{2,1}(\mathrm{A}_{2,1})^2$    & 62 & 384  & $\mathrm{E}_{8,2}\mathrm{B}_{8,1}$ \\
\hline
27 & 96  & $\mathrm{A}_{8,3}(\mathrm{A}_{2,1})^2$               & 63 & 408  & $\mathrm{A}_{15,1}\mathrm{D}_{9,1}$ \\
\hline
28 & 96  & $\mathrm{E}_{6,4}\mathrm{C}_{2,1}\mathrm{A}_{2,1}$            & 64 & 456  & $\mathrm{D}_{10,1}(\mathrm{E}_{7,1})^2$ \\
\hline
29 & 108 & $(\mathrm{B}_{4,2})^3$                      & 65 & 456  & $\mathrm{A}_{17,1}\mathrm{E}_{7,1}$ \\
\hline
30 & 120 & $(\mathrm{A}_{3,1})^8$                      & 66 & 552  & $(\mathrm{D}_{12,1})^2$ \\
\hline
31 & 120 & $(\mathrm{D}_{5,2})^2(\mathrm{A}_{3,1})^2$           & 67 & 624  & $\mathrm{A}_{24,1}$ \\
\hline
32 & 120 & $\mathrm{E}_{6,3}(\mathrm{G}_{2,1})^3$               & 68 & 744  & $(\mathrm{E}_{8,1})^3$ \\
\hline
33 & 120 & $\mathrm{A}_{7,2}(\mathrm{C}_{3,1})^2\mathrm{A}_{3,1}$        & 69 & 744  & $\mathrm{D}_{16,1}\mathrm{E}_{8,1}$ \\
\hline
34 & 120 & $\mathrm{D}_{7,3}\mathrm{A}_{3,1}\mathrm{G}_{2,1}$            & 70 & 1128 & $\mathrm{D}_{24,1}$ \\
\hline
35 & 120 & $\mathrm{C}_{7,2}\mathrm{A}_{3,1}$                   &    &      &  \\
\hline
\end{tabularx}
\end{table}

\section{Summary of 3D TQFTs and Condensable Algebras} \label{GaugingReviewAppendix}

This appendix provides a brief summary of the techniques used throughout the main text to gauge abelian and non-invertible one-form symmetries in 3D TQFTs, a process known also as anyon condensation. To set up the notation we summarize a few concepts from  the algebraic theory of anyons. We follow appendix A of \cite{Cordova:2024goh}, other useful references include \cite{Kitaev:2005hzj, Benini:2018reh,Fuchs:2002cm} and \cite{Kong:2013aya} for a more thorough expositions of anyon condensation. 

A 3D TQFT is defined by a finite set of data. To begin, a TQFT is described in terms of a finite set of elementary topological line operators known as anyons $a$. Any other line operator can be expressed as the direct sum of anyons. A TQFT is also defined in terms of a set of fusion rules, dictating how anyons fuse with each other,
\begin{equation} \label{fusionrules}
    a \times b = \sum_{c} N_{ab}^{c} \, c \, ,
\end{equation}
where the $N_{ab}^{c}$ are non-negative integers that count how many distinct ways anyons $a$ and $b$ can fuse to yield anyon $c$. 
The fusion of anyons is both commutative and associative, and there always exists an identity anyon ``1'' such that $1 \times a = a \times 1 = a$. Furthermore, every anyon $a$ has a unique conjugate that we denote by $\bar{a}$ and defined by the property that $a \times \bar{a} = 1 + \cdots$. An anyon is \textit{invertible} if its fusion with its conjugate yields
only the identity, i.e. $a \times \bar a=1$. Such anyons are also called abelian because invertible anyons form an abelian group under fusion. If $a \times \bar a$ contains
anyons in addition to the identity anyon $a$ is called \textit{non-invertible}.

An important quantity that can be defined from the fusion rules \eqref{fusionrules} is the quantum dimension $d_{a}$ of an anyon $a$, which may be understood as the expectation value of an unknot of $a$:
\begin{equation} \label{QuantumDimension}
    d_{a} = \includegraphics[scale=0.047, valign=c]{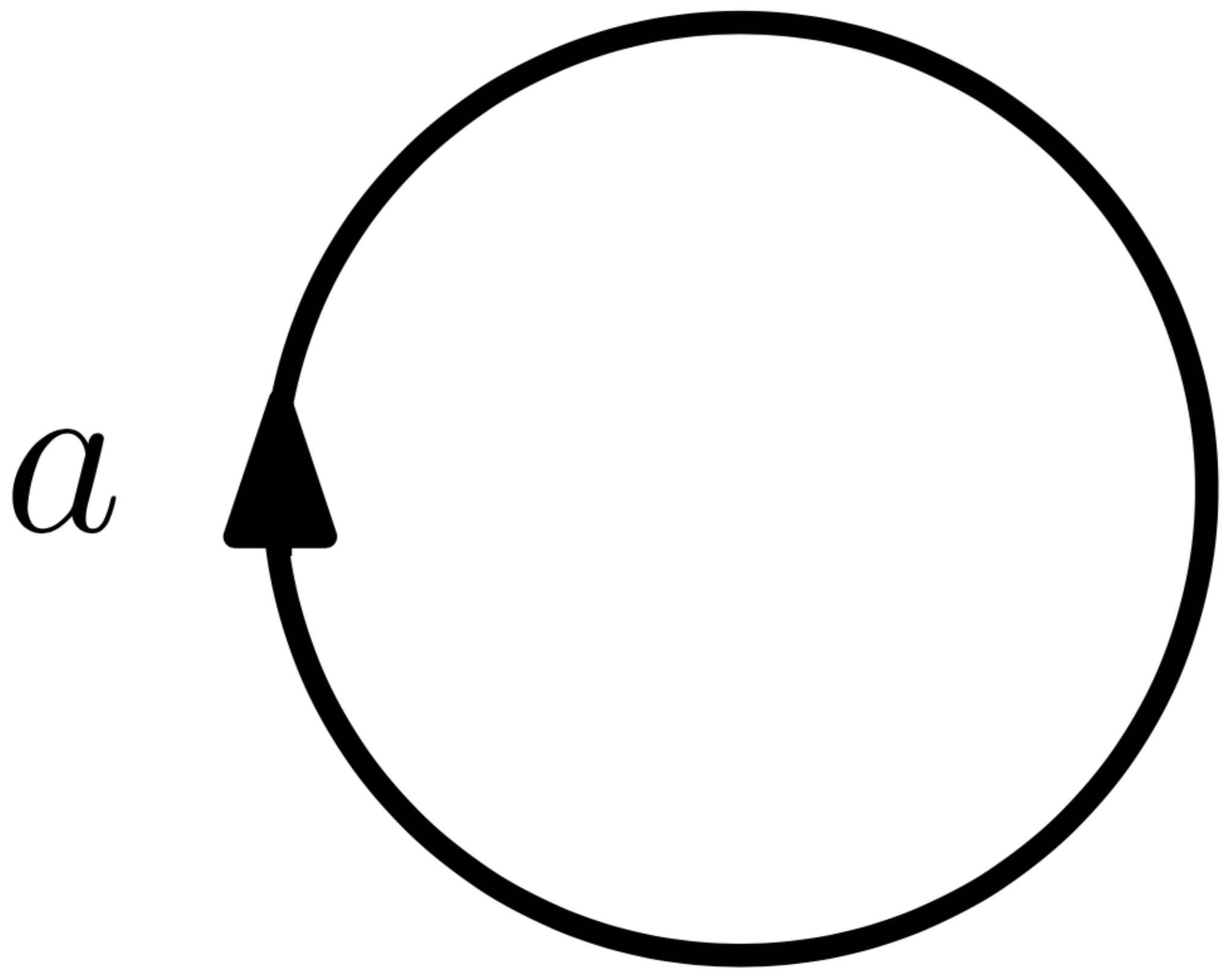} \ .
\end{equation}
The $d_a$ are the unique positive real solutions to
\begin{equation} \label{fusionrulesofqdimensions}
    d_{a} d_{b} = \sum_{c} N_{ab}^{c} \, d_{c} 
\end{equation}
with $d_1=1$ for the identity anyon.
In the correspondence between 3D TQFT and 2D CFT they can also be defined in terms of components of the modular S matrix,
\begin{equation}
d_a= \frac{S_{a0}}{S_{00}} \, ,
\end{equation}
with $0$ denoting the identity primary.
The total quantum dimension of a MTC/3D TQFT $\mathcal{C}$ is defined as
\begin{equation}
    \mathrm{dim}(\mathcal{C}) = \sum_{a \in \mathcal{C}} d_{a}^{2} \, .
\end{equation}

Anyons in a 3D TQFT may braid with each other. This is a property that is summarized by two quantities: the topological spin $\theta_{a}$ of an anyon $a$ and the modular $S$-matrix $S_{ab}$ between anyons $a$ and $b$:
\begin{equation} \label{TopologicalTwist}
    \includegraphics[scale=0.09, valign=c]{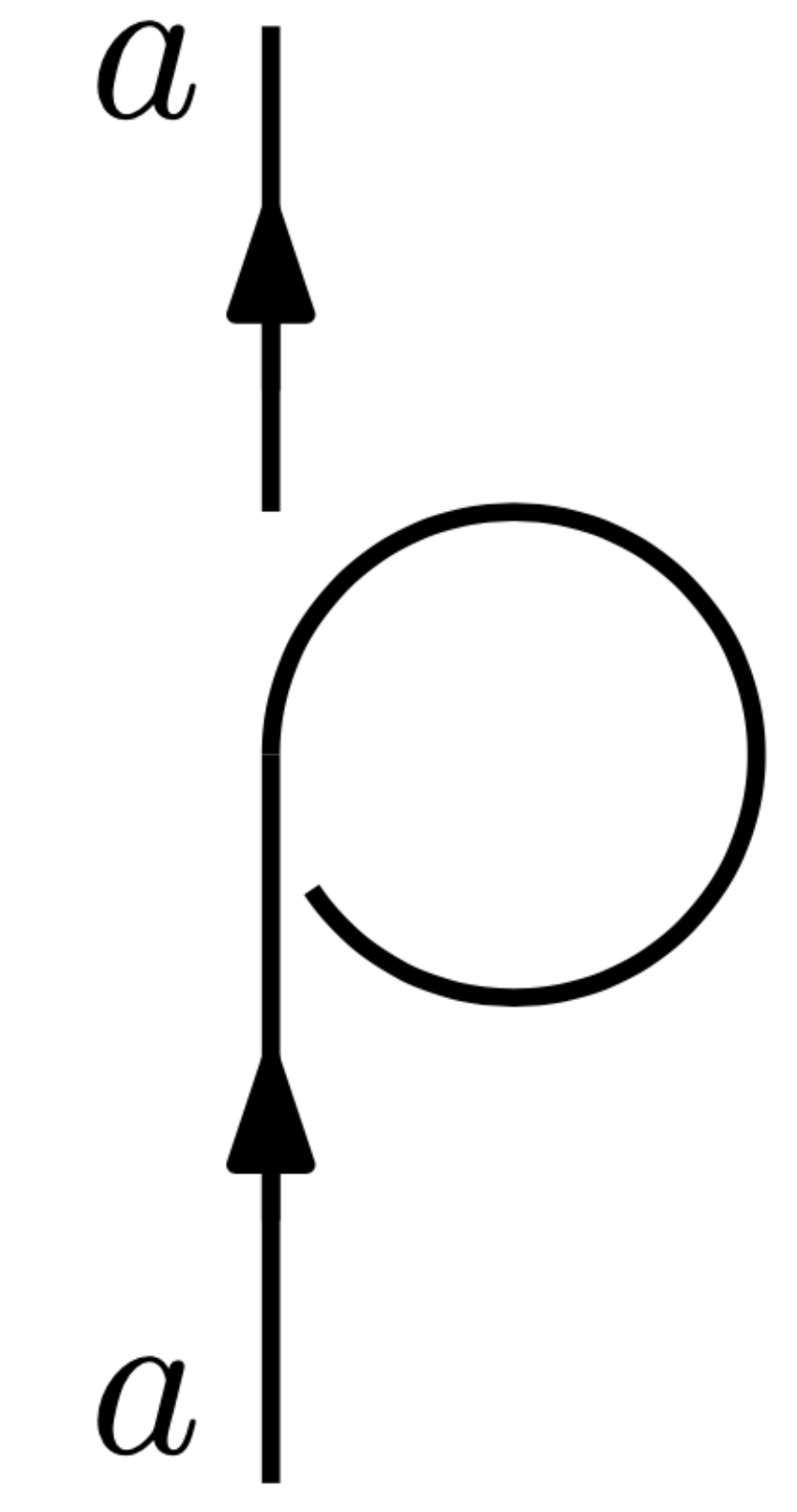} = \theta_{a} \includegraphics[scale=0.09, valign=c]{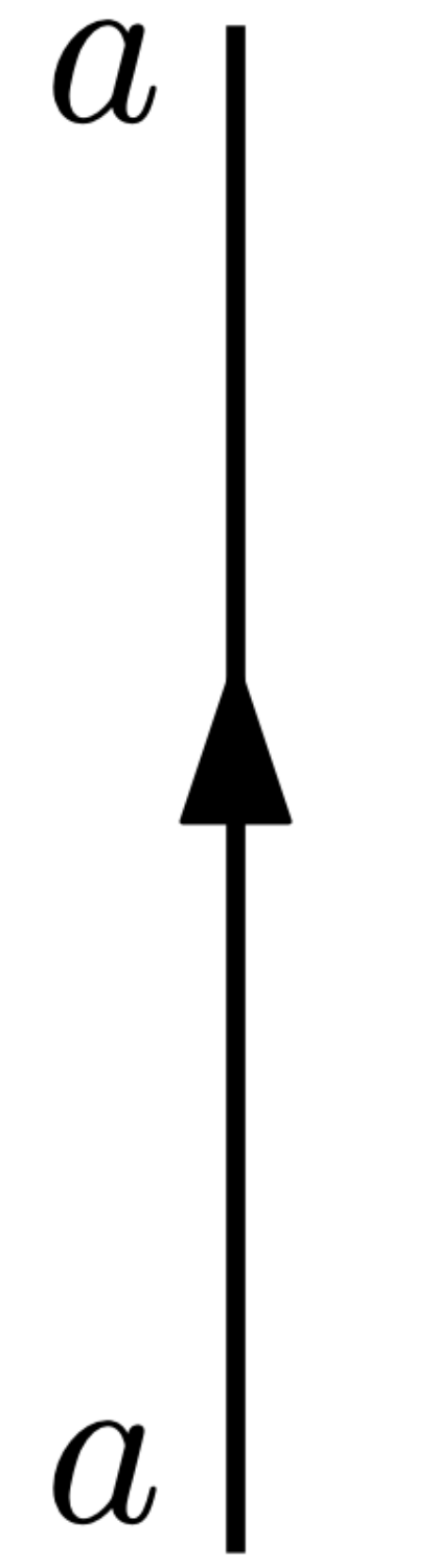}, \quad \quad \quad \quad \includegraphics[scale=0.08, valign=c]{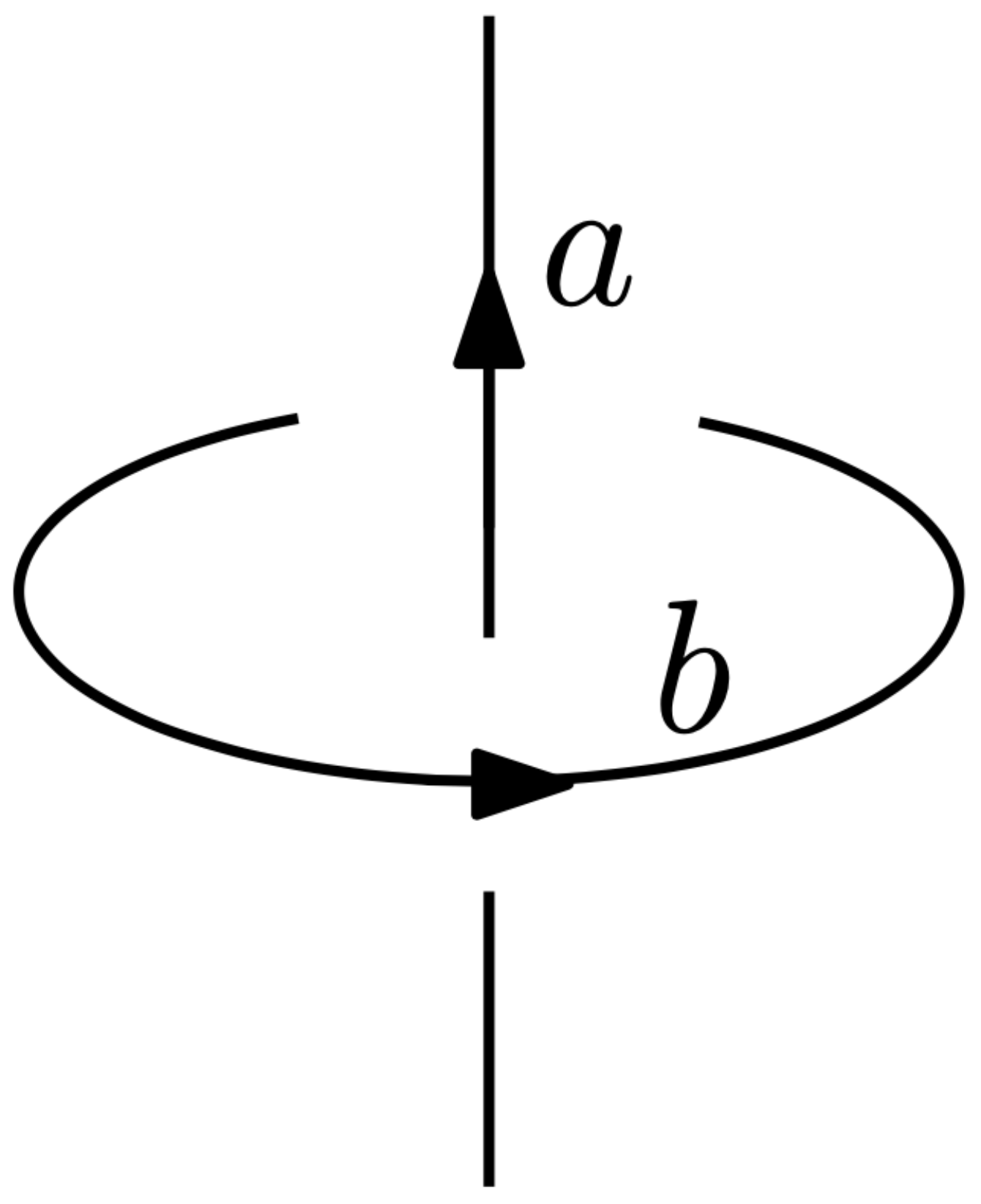} = \frac{S_{ab}}{S_{a0}} \includegraphics[scale=0.08, valign=c]{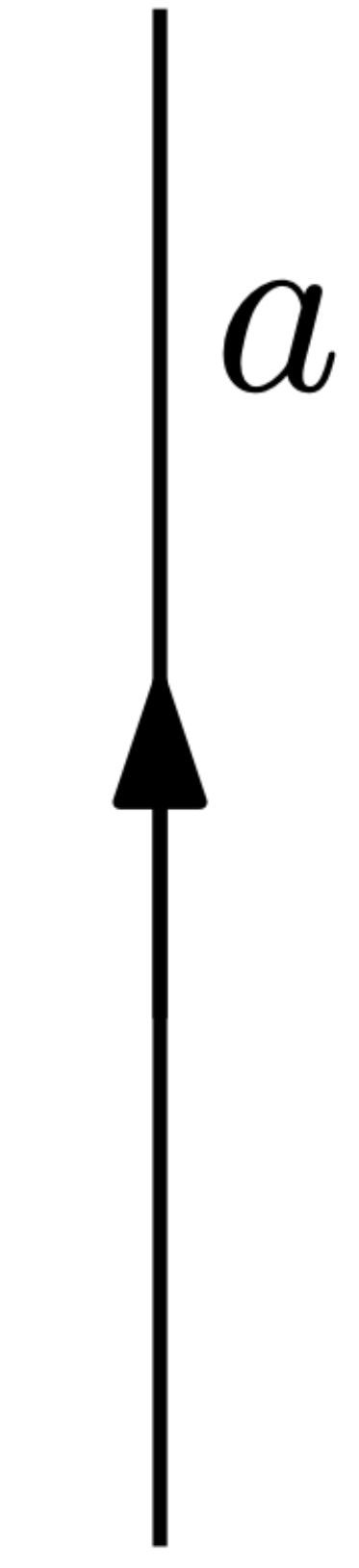}.
\end{equation}
If a correspondence with 2D RCFT data is applicable through bulk-boundary correspondence, the set of anyons $a$ corresponds to the primaries of the RCFT, $\theta_{a} = e^{2 \pi i h_{a}}$, where $h_{a}$ denotes the conformal weight of the respective primary, and $S_{ab}$ coincides with the conventional modular $S$-matrix of RCFT. Often, we provide $h_{a}$ instead of $\theta_{a}$ for the topological spin of a line. If $\theta_{a} = 1$, the corresponding line $a$ is called a \textit{boson}. The ratio $S_{ab}/S_{a0}$ is called the braiding of the anyon $b$ around $a$, and is a phase when $b$ is abelian. 

A full specification of a 3D TQFT also requires specifying the so-called $F$- and $R$-symbols that encode certain associativity and commutativity properties of the anyons. We will not make use of these symbols in our discussion and refer to the aforementioned references for further information.

We now characterize gaugeable symmetries in 3D TQFTs. For an abelian anyon $a$, a necessary and sufficient condition to be gaugeable is that its topological spin is trivial,     
$\theta_a=1$ ($h_a \in \mathbb{Z}$) and that the set of condensing anyons has trivial mutual braiding \cite{Gaiotto:2014kfa}. 

To determine the spectrum of the 3D TQFT after gauging such an abelian symmetry we follow the three-step gauging rule of \cite{Moore:1989yh,Hsin:2018vcg}:
\begin{itemize}
    \item Remove all lines that braid non-trivially with the condensing abelian anyons.  Such removed lines are often said to be \emph{confined}. 
    \item Identify any remaining lines that differ by fusion with the condensing abelian anyons.  This is the step of forming gauge orbits. 
    \item Split any remaining line $a$ that is invariant under fusion with $s$ condensing abelian anyons into $s$ distinct lines in the gauged theory. 
\end{itemize}

In the application of this formalism to verifying Level-Rank duality,  abelian anyon condensation (gauging of an abelian one-form symmetry) reduces to
checking $\theta_a=1$ for the condensing anyons and then applying the three-step procedure above. This algorithm will be our main tool for abelian anyon condensation. After gauging the modular $S$-matrix and fusion rules can be determined as in \cite{Fuchs:1996dd}.

More generally, we must consider gauging non-invertible one-form symmetries, also known as non-abelian anyon condensation. In this case, the object we gauge in a 3D TQFT $\mathcal{C}$ is called a condensable algebra,  also known in the literature as a connected \'etale algebra or a a commutative Frobenius algebra. A condensable algebra is a formal sum of  anyons:
\begin{equation} 
    \mathcal{A} = \sum_{a \in \mathcal{C}} n_{a} \, a \, ,
\end{equation}
where the $n_{a}$ are non-negative integers, and the identity line is always included with multiplicity one:  $1 \in \mathcal{A}$, $n_{1} = 1$. As a simple example, when we gauge abelian bosons that form a group $G$, the corresponding condensable algebra is:
\begin{equation}
    \mathcal{A} = \sum_{g \in G} g \, .
\end{equation}
The TQFT obtained after gauging is denoted as $\mathcal{C}/\mathcal{A}$, or as $\mathcal{C}^{0}_{\mathcal{A}}$, depending on context.

A condensable algebra must satisfy three constraints. 
\begin{itemize}
\item It contains only bosonic lines
\begin{equation}
    a \in \mathcal{A} \Longrightarrow \theta_{a} = 1.
\end{equation}
\item The coefficients $n_a$ are compatible with fusion
\begin{equation}
    n_{a} n_{b} \leq \sum_{c} N^{c}_{ab} n_{c} \, .
\end{equation}
\item
The total quantum dimension of the resulting MTC after condensation is determined by the dimension of the original MTC and the dimension of the condensable algebra as
\begin{equation}  \label{quantumdimensionconstraint}
    \mathrm{dim}(\mathcal{C}/\mathcal{A}) = \mathrm{dim}(\mathcal{C}^{0}_{\mathcal{A}}) = \frac{\mathrm{dim}(\mathcal{C})}{\mathrm{dim}(\mathcal{A})^{2}}
\end{equation}
with
\begin{equation}
    \mathrm{dim}(\mathcal{A}) = \sum_{a} n_{a} \, d_{a}.
\end{equation}
\end{itemize}
In the special case where $\mathrm{dim}(\mathcal{C}) = \mathrm{dim}(\mathcal{A})^{2}$ so that $\mathrm{dim}(\mathcal{C}/\mathcal{A}) = 1$, $\mathcal{A}$ is called a Lagrangian algebra. Gauging $\mathcal{A}$ in $\mathcal{C}$ results in a theory with trivial anyon data, but the same chiral central charge. In other words, a purely gravitational Chern-Simons theory. Gauging a Lagrangian algebra on half of spacetime provides a topological boundary for $\mathcal{C}$.

Strictly speaking, a condensable algebra comes equipped with a definition of a (co)-multiplication map satisfying certain associativity and commutativity properties. Finding such a multiplication map is, however, a daunting task in general. Fulfilling the constraints above is already a highly non-trivial task, so for practical purposes we mainly use these constraints as necessary checks in testing Level-Rank dualities involving non-abelian anyon condensation. For a complete definition of condensable algebras, we refer to the previous references.

Now that we have determined the properties a condensable algebra must fulfill in order to be gauged, we turn to characterizing the resulting TQFT $\mathcal{C}/\mathcal{A}$ after gauging. This can be done as follows. First, the anyons $a$ in $\mathcal{C}$ can be split into line operators of $\mathcal{C}/\mathcal{A}$ according to
\begin{equation} \label{splittings}
    a \longrightarrow \sum_{i} n_{a}^{i} \, a_{i},
\end{equation}
where $n_{a}^{i}$ non-negative integers. Not all the $a_{i}$'s in \eqref{splittings} correspond to independent line operators. The condensation enforces the identification of some of the $a_{i}$'s (as described in detail in the two examples in Section \ref{noninvertiblecondensationexamplessection}). Moreover, in this splitting we have to differentiate genuine, or unconfined, line operators from non-genuine, or confined, line operators. 

A practical way to do this is as follows. We first say that an $a_i$ has a {\it lift} in $\mathcal{C}$ if it appears in the splitting of an $a \in {\cal C}$. If a given $a_{i}$ has a lift to more than one $a \in \mathcal{C}$  and some of the $a$'s have different topological spins, the corresponding $a_{i}$ is confined. For example, one can see from Table \ref{E7lv3table} that the element $(2 \mathbf{w}_6)_1$ in $\mathrm{E}_{7,3}/{\cal A}$ has a lift to both $(2 \mathbf{w}_6)$ and $(\mathbf{w}_2)$ in $\mathrm{E}_{7,3}$ and that these two lines have different topological spins, implying the confinement of $(2 \mathbf{w}_6)_1$. When all the lifts have the same topological spins, we identify the corresponding $a_{i}$ as an unconfined, genuine line operator of $\mathcal{C}/\mathcal{A}$.

The consistency conditions from which we can determine, at least in many practical examples, the splittings and identifications mentioned above are the following. Simple anyons in the condensable algebra $\mathcal{A} = \sum_{a \in \mathcal{C}} n_{a} \, a,$ always have a component of the identity line in the splitting: 
\begin{equation}
    a \in \mathcal{A} \Longrightarrow a \rightarrow \, n_{a} \, 1 + \cdots.
\end{equation}

Moreover, the components $a_{i}$ are constrained to satisfy the following set of consistency conditions: \\
\begin{itemize}
    \item $a \longrightarrow \sum_{i} n_{a}^{i} \, a_{i} \Longrightarrow d_{a} \longrightarrow \sum_{i} n_{a}^{i} \, d_{a_{i}}$ 
    \item $a \longrightarrow \sum_{i} n_{a}^{i} \, a_{i} \Longrightarrow  \bar{a} \longrightarrow \sum_{i} n_{a}^{i} \, \bar{a}_{i}.$ 
    \item $a \times b = \sum_{c}  N_{ab}^{c} \, c \Longrightarrow \Big( \sum_{i} n_{a}^{i} \, a_{i} \Big) \times \Big( \sum_{j} n_{b}^{j} \, b_{j} \Big) = \sum_{k} \sum_{c} N_{ab}^{c} n_{c}^{k} \, c_{k}.$ 
\end{itemize}
We assume as well that when we perform non-abelian anyon condensation, the fusion category of unconfined excitations as well as that of all line operators, both confined plus unconfined, satisfies the standard conditions of associativity, existence of a unique identity line, and existence of unique conjugate pairs with a unique way to fuse to the identity line. 

\section{Summary of Anyon Data} \label{SummaryAnyonDataAppendix}

For quick reference, in this appendix we summarize some anyon data that appears multiple times throughout the main text.

\subsection*{$\mathrm{SU}(N)_{1}$}

The first MTC that we summarize is $\mathrm{SU}(N)_{1}$. The theory consists of $N$ lines with conformal weights/topological spins:
\begin{equation}
    h_{i} = \frac{i(N-i)}{2N}, \quad i = 0, \ldots, N-1.
\end{equation}
The lines fulfill $\mathbb{Z}_{N}$ fusion rules: $i \times j = i+j \mod N$, while the modular $S$-matrix is given by:
\begin{equation}
    S_{ij} = \frac{1}{\sqrt{N}} \omega^{ij}.
\end{equation}
where $\omega = e^{2 \pi i / N}$. Clearly, from this expression we see all lines have unit quantum dimension.

\subsection*{$\mathrm{Spin}(\nu)_{1}$}

The MTC $\mathrm{Spin}(\nu)_{1}$ for odd $\nu$ always consists of three lines $1,v,\sigma$ with conformal weights/topological spins:
\begin{equation}
    h_{1} = 0, \quad h_{v} = 1/2, \quad h_{\sigma} = \nu/16,
\end{equation}
fusion rules:
\begin{equation}
    v \times v = 1, \quad v \times \sigma = \sigma \times v = \sigma, \quad \sigma \times \sigma = 1 + v.
\end{equation}
and quantum dimensions
\begin{equation}
    d_{1} = 1, \quad d_{v} = 1, \quad d_{\sigma} = \sqrt{2}.
\end{equation}

\subsection*{$\mathrm{G}_{2,1} \ \& \ \mathrm{F}_{4,1}$}

Both the $\mathrm{G}_{2,1}$ and $\mathrm{F}_{4,1}$ MTCs have only one non-trivial line $\tau$ of quantum dimension $\phi = (1 + \sqrt{5})/2$ and topological spin:
\begin{equation}
    h_{\tau} = \frac{2}{5}, \, \frac{3}{5}, 
\end{equation}
for $\mathrm{G}_{2,1}$ and $\mathrm{F}_{4,1}$ respectively. The fusion rule is
\begin{equation}
    \tau \times \tau = 1 + \tau.
\end{equation}
Notice that $\mathrm{G}_{2,1}$ and $\mathrm{F}_{4,1}$ are Level-Rank dual to each other \cite{Cordova:2018qvg}. More precisely:
\begin{equation}
    \mathrm{G}_{2,1} \cong \mathrm{F}_{4,-1} - 16 \CSgrav.
\end{equation}

\section{Time-Reversal Invariance of Bosonic $\mathrm{SU}(k)_{1}$} \label{SU(k)1TimeReversalAppendix}

In this appendix we show for completeness that, as a bosonic theory, the $\mathrm{SU}(k)_{1}$ TQFT for $k$ odd is time-reversal invariant if and only if $-1$ is a quadratic residue modulo $k$. See \cite{Geiko:2022qjy, Delmastro:2019vnj} for other works studying time-reversal of abelian Chern-Simons theories.

First, recall that an abelian TQFT enjoys time-reversal symmetry if there exists a permutation $P$ of the anyons $i$ that preserves the fusion rules: $i \times j = k \rightarrow P(i) \times P(j) = P(k)$ and such that the topological spin of the anyons are reversed: $\theta_{P(i)} = \theta_{i}^{*}$. If $\mathrm{SU}(k)_{1}$ is time-reversal invariant, it exists $q$ such that:
\begin{equation}
    \frac{q(k-q)}{2k} = -\frac{(k-1)}{2k} \mod 1,
\end{equation}
or equivalently
\begin{equation}
    q(k-q) + k-1 = z2k,
\end{equation}
for some integer $z$. Rearranging, we find:
\begin{equation}
    (q + 1 - 2z)k - q^{2} = 1,
\end{equation}
which is the condition for $-1$ to be a quadratic residue modulo $k$. The other way around, if we assume $-1$ to be a quadratic residue modulo $k$, we have integers $p,q$ such that $pk - q^{2} = 1$. Notice that since $k$ is odd, $p$ and $q$ must have different parities. Given these integers, we can define:
\begin{equation}
    z \coloneqq \frac{(q-p)+1}{2},
\end{equation}
which is an integer since $q$ and $p$ have different parities. Then, we have that
\begin{equation}
    (q + 1 - 2z)k - q^{2} = 1 \Longleftrightarrow q(k-q) + (k-1) = z2k,
\end{equation}
and we deduce the desired property:
\begin{equation}
    \frac{q(k-q)}{2k} = -\frac{(k-1)}{2k} \mod 1,
\end{equation}
which is the condition for the generator of $\mathbb{Z}_{k}$ to be mapped to another anyon with opposite spin. Once we have this, it is straightforward to verify that the rest of the lines permute accordingly.

\bibliographystyle{JHEP}
\bibliography{main}

\end{document}